\documentclass[aps,prb,twocolumn,groupedaddress,floatfix,showpacs]{revtex4-2}

\usepackage{graphicx}
\usepackage{amsmath}
\usepackage{subcaption}
\usepackage{color}
\newcommand{\parl}{\parallel}

\begin{document}


\title{Nonlinear optical properties and Kerr nonlinearity of Rydberg excitons\\ in Cu$_2$O quantum wells}


\author{David Ziemkiewicz}
\email{david.ziemkiewicz@utp.edu.pl}

\author{Gerard Czajkowski}
\author{Karol Karpi\'{n}ski}
\author{Sylwia Zieli\'{n}ska-Raczy\'{n}ska}

 \affiliation{Institute of
Mathematics and Physics, Technical University of Bydgoszcz,
\\ Al. Prof. S. Kaliskiego 7, 85-789 Bydgoszcz, Poland}


\date{\today}

\begin{abstract} 
The quantum confiment of Rydberg excitons (REs) in quantum structures opens the way towards considering  nonlinear interactions in such  systems. We present a theoretical calculation of  optical functions  in the case of a nonlinear coupling between REs in a quantum well with an electromagnetic wave. Using the Real Density Matrix Approach (RDMA), the analytical expressions for a linear and
nonlinear absorption are derived and numerical calculations for Cu$_2$0  quantum wells are performed. The results indicate the conditions in which quantum well confinement states can be observed in linear and nonlinear optical spectra. The Kerr nonlinearity and self-phase modulation in such a system are studied. The effect of Rydberg blockade and the associated optical bleaching are also discussed and confronted with available experimental data.
\end{abstract}


\maketitle
\section{Introduction}
Rydberg physics in semiconductors has started in 2014 by an observation of  highly excited excitonic states with principal quantum numbers as high as $n$=25  in cuprous oxide, the material of a very large exciton binding energy \cite{Kazimierczuk2014}. This experiment revealed a plethora of Rydberg excitons' unusual properties such as extraordinary large dimensions up to 1 $\mu$m, long life-times of order of nanosecond, vulnerability  to interactions with external fields and restrictions of their coupling arising from Rydberg blockade, which precludes a simultaneous  excitation of two Rydberg excitons that are separated by less then a blockade radius $r_b$. 
A lot of papers have been devoted to studies of spectroscopic characteristic of REs in natural and synthetic bulk systems of Cu$_2$O \cite{Heckotter2017, Assman2020, Lynch2021}
(see more references therein). Simultaneously, the explorations of RE   in the field of  quantum optics  have begun by  demonstration of  a generation and control of strong excitonic interactions with the help of two-color pump-probe technique \cite{Heckotter2021},  Rydberg exciton-assisted coupling between microwave and optical fields  \cite{Liam2022} and the experimental verification of   the strong coupling of REs to cavity photons \cite{Orfanakis2022}.  
Moreover, some efforts have been made to investigate  nonlinear interactions of REs with electromagnetic fields \cite{Raczynska2019, Walther2020}. The recent  one-photon experiment has shown a giant nonlinear optical index in a bulk Cu$_2$O crystal,  caused by  sharp Rydberg resonances   and revealed  a Kerr phase-shift much larger than in typical nonlinear crystals \cite{ThomasArxiv2022}.  
Interesting, giant microscopic dimensions of Rydberg excitons together with an intrinsic  Rydberg blockade effect in cuprous oxide cause enhanced nonlinearities at much smaller densities compared with other semiconductors \cite {Kazimierczuk2014, Walther2018}. 

Those results indicate that Rydberg excitons are a unique platform for obtaining strong interactions in  solid systems and allow one to hope for  a realization, in a close future, of  solid state masers \cite{Ziemkiewicz2018,Ziemkiewicz2019} and few-photon devices. The first step to achieve a scalable solid-state platform characterized by controlled interactions between Rydberg excitons and photons to realize such technologically demanding miniaturized systems, consisted in an  investigation of REs'  properties in strongly confined systems such as quantum dots, wires or wells \cite{ Konzelmann2019, Czajkowski2020,Czajkowski2021,Orfanakis2021}.  The experiment, which has verified a change of oscillator strength due to quantum confinement of REs in a nanoscale system \cite{Orfanakis2021}, is an important step towards exploiting their large nonlinearities for quantum applications. The recent progress in fabricating synthetic cuprous oxide elements has shown an enormous progress of their quality manifested by observations of high excitonic states \cite{Lynch2021, Takahata, Steinhauer} and now the natural direction of subsequent explorations seems to be the study of a nonlinear interaction between confined REs and light.
A great challenge in quantum optics is an accomplishment  of
 the gigant Kerr nonlinearities in solid-state low-dimensional media. This  phenomenon was realized in semiconductor quantum wells mostly under the conditions of the ellectromagnetically induced transparency \cite{Hamedi, Kosionis, Zhu} or in ultrathin gold films \cite{Qian}.  In our paper we propose a realization of the Kerr nonlinearity in the  Cu$_2$O quantum well with REs, taking advantage from the fact that confinement effects result in a significant optical Kerr susceptibility.

The theoretical tool which we use  to calculate the optical functions for nonlinear interacion of electromagnetic radiation with Rydberg excitons in a quantum well is a mesoscopic method, called Real Density Matrix Approach \cite{StB87,CBass}. It allows for a calculation of transition probability amplitudes taking into account finite lifetimes, all mutual and external interactions and a system geometry. The detailed description  of RDMA as well as a presentation of the iteration procedure, which allows one to obtain a nonlinear susceptibility for Rydberg excitons confined in a quantum well, is presented in Sections II and III.  The effect of the Rydberg blockade is also included into our treatment and is considered in Sec. IV. Phase-sensitive Kerr nonlinearity appearing in discussed system is examined in Sec. V.  Sec. VI contains the presentation of numerical results and their discussion, while the summary and conclusions of our paper are presented in the last Sec.VII. 

\section{Real Density Matrix Approach}
\subsection{Basic equations}
Our discussion follows the scheme of
Refs.\cite{Raczynska2019,ThomasArxiv2022} adapted for the case of a quantum
well. In the RDMA approach that nonlinear response will be
described by a set of three coupled  constitutive
equations:  for the coherent amplitude $Y({\bf r}_1, {\bf r}_2)$
representing the exciton density,  for the density matrix $C({\bf r}_1, {\bf r}_2)$ for
electrons  (assuming a non-degenerate
conduction band), and  for the density matrix for the holes $ D({\bf r}_1, {\bf r}_2)$ in
the valence band. Denoting  $
 Y({\bf r}_1, {\bf r}_2)=Y_{12}$,
\noindent the constitutive equations  take the following form \cite{Czajkowski2021}\\
- for the coherent amplitude
\begin{eqnarray}\label{interband1}
 & &{
i}\hbar\partial_tY_{12}-H_{\hbox{\tiny QW}}Y_{12}=-{\bf M}{\bf
E}({\bf R}_{12})\nonumber
\\
& &+{\bf E}_1{\bf M}_{0}C_{12}+{\bf E}_2{\bf M}_{0}D_{12}+{
i}\hbar\left(\frac{\partial Y_{12}}{\partial t}\right)_{{\rm
irrev}},
\end{eqnarray}
- for the conduction band \begin{eqnarray}
& &\label{conduction1}{i}\hbar\partial_tC_{12}+H_{ee}C_{12}={\bf M}_{0}({\bf E}_1 Y_{12}-{\bf E}_2Y^*_{21})\nonumber\\
& &+{i}\hbar\left(\frac{\partial C_{12}}{\partial t}\right)_{{\rm
irrev}},\end{eqnarray} 
- for the valence band 
\begin{eqnarray} &
&\label{valence1}{i}\hbar\partial_tD_{21}-H_{hh}D_{21}={\bf
M}_{0}({\bf E}_2 Y_{12}-{\bf E}_1Y^*_{21})\nonumber\\
&&+ {i}\hbar\left(\frac{\partial D_{21}}{\partial t}\right)_{{\rm
irrev}},
\end{eqnarray}
\noindent where the operator $H_{\hbox{\tiny QW}}$ is the quantum
well Hamiltonian, which includes the terms $V_e,V_h$ related to the
electron and hole confinement and the mutual Coulomb interaction $V_{eh}$
\begin{eqnarray}\label{hamiltonianeh}
&&H_{\hbox{\tiny
QW}}=E_g-\frac{\hbar^2}{2m_e}\partial_{z_e}^2-\frac{\hbar^2}{2m_h}\partial_{z_h}^2
-\frac{\hbar^2}{2M_{\hbox{\tiny tot}}}\hbox{\boldmath$\nabla$}_{R_\parl}^2\nonumber\\
&&-\frac{\hbar^2}{2\mu}\hbox{\boldmath$\nabla$}^2_\rho+V_{e}(z_e)+V_h(z_h)+V_{eh},
\end{eqnarray}
\noindent with the separation of the center-of-mass coordinate
${\bf R}_\parl$ from the relative coordinate
$\hbox{\boldmath$\rho$}$ on the plane $(x,y)$, e.g. $\boldmath\rho = (r_1-r_2)_\parallel$ and
\begin{eqnarray}
&
&H_{ee}=-\frac{\hbar^2}{2m_e}(\hbox{\boldmath$\nabla$}_1^2-\hbox{\boldmath$\nabla$}_2^2),\nonumber
\\
&
&H_{hh}=-\frac{\hbar^2}{2m_{h}}(\hbox{\boldmath$\nabla$}_1^2-\hbox{\boldmath$\nabla$}_2^2),
\end{eqnarray}
and ${\bf E_1}={\bf E(r_1)}$, ${\bf E_2}={\bf E(r_2)}$. In the case of a quantum well with thickness that is significantly smaller than the light wavelength, one can assume uniform field ${\bf E_1}={\bf E_2}={\bf E}$. The center of mass coordinate is
\begin{equation}\label{com}
{\bf R}={\bf R}_{12}=\frac{m_h{\bf r}_1+m_e{\bf r}_2}{m_h+m_e}.
\end{equation}
\noindent In the above formulas $m_e, m_h$ are the electron and
the hole effective masses (the effective mass
tensors in general), $M_{tot}$ is the total exciton mass and $\mu$ the
 reduced mass of electron-hole pair. The smeared-out transition
dipole density ${\bf M}({\bf r}$) is related to the bilocality of
the amplitude $Y_{12}$ and describes the quantum coherence between the
macroscopic electromagnetic field and the inter-band transitions
(see, for example, Refs. \cite{StB87,CBass}); the detailed derivation of ${\bf M}({\bf r})$ is described in Ref. \cite{PRB93}. We assume that the
carrier motion in the $z$-direction is governed by the no-escape
boundary conditions.
 With this assumptions, the QW Hamiltonian has the
form
\begin{eqnarray}\label{QWparabHamilt1}
&&H_{\hbox{\tiny
QW}}=E_g+\frac{p_{z_e}^2}{2m_e}+\frac{p_{z_h}^2}{2m_h}+V(z_e)+V(z_h)\nonumber\\
&&+ H_{\hbox{\tiny Coul}}^{(2D)}(\hbox{\boldmath$\rho$}),
\end{eqnarray}
\noindent where
\begin{eqnarray}\label{1dimoschamiltonian1}
&&V(z_{e,h})=0\qquad\hbox{for}\quad 0\leq z\leq L,\nonumber\\
&&V(z_{e,h})=\infty\qquad\hbox{for}\quad z<0,\quad z>L,
\end{eqnarray}
 $H_{\hbox{\tiny Coul}}^{(2D)}$ is the
two-dimensional Coulomb Hamiltonian 
\begin{equation}\label{2dimCoulombhamilt1}
H_{\hbox{\tiny Coul}}^{(2D)}(\hbox{\boldmath$\rho$})=\frac{{\bf
p}^2_{\parallel}}{2\mu_{\parallel}}-\frac{e^2}{4\pi\epsilon_0\epsilon_b\rho}.
\end{equation}
\noindent We consider here the strong confinement regime, where the confinement energy exceeds the Coulomb energy. The resulting
coherent amplitude $Y_{12}$ determines the excitonic part of the
polarization of the medium
\begin{eqnarray}\label{polarization1}
&&{\bf P}({\bf R},t)=2\int{\rm d}^3r\,\textbf{M}^*({\bf
r})\hbox{Re}~Y_{12}({\bf
R},{\bf r},t)\nonumber\\
&&=\int{\rm d}^3r\textbf{M}^*({\bf r})[Y_{12}({\bf R},{\bf
r},t)+\hbox{c.c}],
\end{eqnarray}
\noindent where ${\bf r}={\bf r}_1-{\bf r}_2$ is the electron-hole
relative coordinate. 
The linear optical functions are obtained by solving the
interband equation (\ref{interband1}) together with the
corresponding Maxwell equation, where the polarization
(\ref{polarization1}) acts as a source.  Using the entire set of constitutive
equations (\ref{interband1})-(\ref{valence1}) one can  the nonlinear optical functions. While  a
general solution of this problem  seems to be inaccessible, but 
in specific situations such a solution can be found, i.e., if one
assumes that the matrices $Y, C$ and $D$ can be expanded in powers
of the electric field ${\bf E}$, an iteration scheme can be used.

The relevant expansion of the polarization in powers of the field
has the form
\begin{equation}
P({\bf k},\omega)=\epsilon_0 E({\bf k},\omega) \left[ \chi^{(1)}+
\chi^{(3)}(\omega,-\omega,\omega)\vert E({\bf
k},\omega)\vert^2+\ldots\right],
\end{equation}
\noindent where $\chi^{(1)}$  and $\chi^{(3)}$ are the linear and the
nonlinear parts of the susceptibility. Although the above equations apparently resemble those describing the nonlinear case of the bulk crystal with Rydberg excitons \cite{Raczynska2019} we present full theoretical approach for the sake of completeness and it should be stressed that taking into account the confinement  interaction significantly changes the results.
\subsection{Iteration}
 We calculate the QW optical functions iteratively from the constitutive  equations
(\ref{interband1})-(\ref{valence1}). The first step in the
iteration consists of solving the equation (\ref{interband1})( skipping the second and the third terms in its r.h.s.)
which we take in the form
\begin{equation}\label{Ylin}
{i}\hbar\partial_tY^{(1)}_{12}-H_{\hbox{\tiny
QW}}Y^{(1)}_{12}=-{\bf M}{\bf E}+{i}\hbar\left(\frac{\partial
Y_{12}^{(1)}}{\partial t}\right)_{\hbox{\tiny irrev}}.
\end{equation}
It should be mentioned that  we use  the long-wave approximation, which allows to neglect the spatial distribution of the electromagnetic wave inside the quantum well.

For the irreversible part, assuming a relaxation time approximation one gets
\begin{equation}
\left(\frac{\partial Y_{12}^{(1)}}{\partial t}\right)_{\hbox{\tiny
irrev}}=-\frac{1}{T_{2}} Y_{12} = \frac{-\Gamma}{\hbar}Y_{12}.
\end{equation}
with ${\mit\Gamma}=\hbar/T_{2}$ being a dissipation constant. Considering  nonlinear effects the non-resonant parts of the coherent amplitude $Y$ have to be taken into account; so for
the electric field ${\bf E}$ in the medium of the form
\begin{equation}
{\bf E}={\bf E}({\bf R},t)+{\bf E}^*({\bf R},t)= {\bf
E}_0e^{{ i}({\bf kR}-\omega t)} + {\bf E}_0e^{-{ i}({\bf
kR}-\omega t)},
\end{equation}
\noindent  equation (\ref{Ylin}) generates two
equations: one for an amplitude $Y_{-}^{(1)}\,\propto \exp(-{
i}\omega t)$, and the second for the non-resonant part
$Y^{(1)}_{+}\,\propto \exp({ i}\omega t)$,
\begin{eqnarray}\label{linear}
& &{i}\hbar\left( i\omega+\frac{1}{T_{2}}\right)Y^{(1)}_{12+}-
H_{eh}Y^{(1)}_{12+}=-{\bf M}{\bf E}^*({\bf R},t),\nonumber
\\
&&\\ & &{i}\hbar\left(-{
i}\omega+\frac{1}{T_{2}}\right)Y^{(1)}_{12-}- H_{\hbox{\tiny
QW}}Y^{(1)}_{12-}=-{\bf M}{\bf E}({\bf R},t).\nonumber
\end{eqnarray}
In what follows we consider only one component of both ${\bf E}$ and
${\bf M}$. Similarly as in Ref.~\cite{Raczynska2019},
we look for the solution in terms of eigenfunctions of the
Hamiltonian $H_{\hbox{\tiny QW}}$, which now contains the confinement terms, so these eigenfunctions have the following form 
\begin{equation}\label{QWeigen}
\Psi_{jmN_eN_h}(r,\phi,z_e,z_h)=\psi_{jm}(r,\phi)
\psi^{(1D)}_{L,N_e}(z_e)\psi^{(1D)}_{L,N_h}(z_h)
\end{equation}

 where $\textbf{r}=(r,\phi)$ is the two-dimensional space vector,  $\psi_{jm}$ are the normalized eigenfunctions of the
2-dimensional Coulomb Hamiltonian,
\begin{eqnarray}\label{2_Dim_Eigen}
 &&\psi_{jm}(r,\phi)=R_{jm}(r)\frac{e^{im\phi}}{\sqrt{2\pi}},\nonumber\\
&&R_{jm}=C_{jm}\left(4\kappa\frac{r}{a^*}\right)^m
e^{-2\kappa_{jm}r/a^*}\,M\left(-j,2\vert
m\vert+1,4\kappa_{jm}\frac{r}{a^*}\right),\nonumber\\
&&\kappa_{jm}=\frac{1}{1+2(j+\vert m\vert)},\\
&&C_{jm}=\frac{1}{a^*}4\kappa_{jm}^{3/2}\frac{1}{(2m)!}\frac{[(j+2m)!]^{1/2}}{[j!]^{1/2}},\nonumber
\end{eqnarray}
 with  the Kummer function \cite{Abramovitz}
$M(a,b,z)$ (the confluent hypergeometric function),
 and
$\psi^{(1D)}_{\alpha,N}(z)$ (N=0,1,...) are the quantum oscillator
eigenfunctions of the Hamiltonian (\ref{1dimoschamiltonian1})
\begin{eqnarray*}\label{eigenf1doscillator}
&&\psi_{L,N_e}^{(1D)}(z_e)=\sqrt{\frac{2}{L}}\cos\left[(2N_e-1)\pi\frac{z_e}{L}\right],\nonumber\\
&&\psi_{L,N_h}^{(1D)}(z_h)=\sqrt{\frac{2}{L}}\cos\left[(2N_h-1)\pi\frac{z_h}{L}\right].\\
\end{eqnarray*}
\noindent
The role of the amplitude $Y_{12}^{(1)}$ obtained in such a way  is two-fold. First, 
substituted into Eq. (\ref{polarization1}) gives the linear
excitonic polarization $P^{(1)}$ and from the relation
$P^{(1)}=\epsilon_0\chi^{(1)}E$ we can calculate the mean
effective linear susceptibility, which is given by the following expression
\begin{eqnarray}\label{chilin1}
& &\chi^{(1)}\left(\omega\right)=\nonumber\\ &&=\chi^{(1)}_0
 \left(\frac{a^*}{L}\right)\,\sum\limits_{0}^{N_{\hbox{\tiny max}}}\sum\limits_{j=0}^J\frac{
 f_{j}^{(2D)}
E_{TjN}}{E_{TjN}^2-(\hbar\omega+i{\mit\Gamma}_j)^2},
\end{eqnarray}
where the summation is  over the confinement state number $N$ and excitonic state number $j$, where $j=0$ is the lowest excitonic state. The oscillator strengths $f_j^{(2D)}$ are given by 
\begin{eqnarray}\label{definitions_chifj}
&&f_{j}^{(2D)}
=48\frac{(j+1)(j+2)}{\left(j+3/2\right)^5}\frac{1}{(1+2\kappa_{j}\rho_0)^8}\nonumber\\
&&\times\left[F\left(-j,4;3;\frac{4\kappa_{j}\rho_0}{1+2\kappa_{j}\rho_0}\right)\right]^2,\\
&&\kappa_{j}=\frac{1}{2j+3},\nonumber
\end{eqnarray}
and the energy terms, including exciton binding energy $E_j$ and quantum well contribution $W_N$ are as follows
\begin{eqnarray}\label{ete}
&&E_{TjN}=E_g+W_N+E_j,\nonumber\\
&&W_N=\left(\frac{N\pi a^*}{L}\right)^2R^*,\qquad N=1,2,\ldots,\\
&&E_j=-\frac{1}{(j+1-\delta)^2}R^*,\quad j =0,1,2,\ldots\nonumber
\end{eqnarray}
$a^*$ is the effective exciton Bohr radius, $R^*$ the exciton
Rydberg energy, $\rho_0=r_0/a^*$ defines the coherence radius, and
$F(a,b;c,z)$ is the hypergeometric series \cite{Abramovitz}. The $\delta=0.2$ is the so-called quantum defect \cite{Kazimierczuk2014}; it should be mentioned that while the most common value of $\delta$ is used here, smaller ones provide a better fit to many experimental results, especially at elevated temperatures \cite{Kang}.
The constant factor $\chi^{(1)}_0$ has the form 
\begin{equation}\label{def_chi1}
\chi_0^{(1)}= \epsilon_b e^{-4\rho_0}\Delta_{LT},
\end{equation}
For simplicity, we can use only one confinement state number by  considering only the largest contribution from $N_e=N_h=N$. Due to the long wave approximation the validity of our considerations is limited regarding the quantum well width $L$, which in turn entails the restriction of the highest observable confinement states $N_{max}$. Specifically, in the case of a thin quantum well, the considerable confinement energy $W_N$ means that for higher $N$ and $j$, the total energy $E_{TjN}$ approaches the band gap, where higher absorption precludes the observation of confinement states.

\section{Iteration procedure: second step}\label{iteration2nd}
  Again, in order to present the detailed derivation of nonlinear susceptibility for a quantum well with Rydberg excitons we recall the procedure in general similar to that presented in \cite{Raczynska2019}, but considering here the low dimensional systems significantly changes the final results. 
 Let us first consider a wave linearly polarized in the $z$
direction. Then $Y_{\pm}^{(1)}$ (\ref{linear}) are inserted into the source terms of the
conduction-band and valence band equations (\ref{conduction1} -
\ref{valence1}).  Solving for stationary solutions and making 
the long wave approximation, we obtain for  both source terms
\begin{eqnarray}
&
&J_{C}=M_{0}\rho_0\left(E_1Y_{12}^{(1)}-E_2Y_{21}^{(1)*}\right)\nonumber
\\
&
&=\frac{2{i}M_{0}\rho_0{E}_0^2}{\hbar}\left[\,\hbox{Im}\,g(-\omega,
{\bf r})+\,\hbox{Im}\,g(\omega, {\bf r})\right]=J_V,
\end{eqnarray}
\noindent where\begin{equation} g(\pm\omega, {\bf
r})=\sum_{j}\frac{c_{jmN_eN_h}\Phi_{jmN_eN_h}({\bf
r})}{\Omega_{jmN_eN_h}\mp\omega-{i}/T_{2jm}}.
\end{equation}
 If
irreversible terms are well defined, the equations
(\ref{valence1}) can be solved and their solutions are then used
in the saturating terms on the r.h.s. of the equations
(\ref{interband1}). Again as in the previous section, we will use a relaxation time approximation and the equations for the matrices $C$  and $D$ are as follows
\begin{eqnarray}\label{relaxation_time}
& &\left(\frac{\partial C}{\partial t}\right)_{{\rm
irrev}}\nonumber\\
&& =-\frac{1}{\tau}\left[C({\bf X},{\bf r},t)-f_{0e}({\bf
r})C({\bf X}, {\bf
r}=\textbf{r}_0,t)\right]-\frac{C(r_0)}{T_1},\nonumber
\\
&&\\
 & &\left(\frac{\partial D}{\partial t}\right)_{{\rm
irrev}}\nonumber\\
&&= -\frac{1}{\tau}\left[D({\bf X},{\bf r},t)-f_{0h}({\bf
r})D({\bf X}, {\bf
r}=\textbf{r}_0,t)\right]-\frac{D(r_0)}{T_1},\nonumber
\end{eqnarray}
\noindent where
\begin{equation}
{\bf X}=\frac{1}{2}\left({\bf r}_e+{\bf r}_h\right),
\end{equation}
\noindent and $f_{0e}, f_{0h}$ are normalized Boltzmann
distributions for electrons and holes, respectively. The relaxation parameter $T_1$
 is due to interband recombination \cite{FrankStahl} and $\tau=1/\Gamma_j << T_1$ is the lifetime corresponding to radiative recombination.
The functions $C,D$ must have the same $p$-symmetry as the
amplitudes $Y$. Thus we use the transport current density
\begin{equation}
\textbf{j}_n(\textbf{r})=\frac{i\hbar}{2m_e}\left(\hbox{\boldmath$\nabla$}_1-\hbox{\boldmath$\nabla$}_2\right)\vert_{\hbox{\tiny
\textbf{r}}_1=\hbox{\tiny \textbf{r}}_2=\textbf{r}},
\end{equation}
and take the $x$-component, which leads to the following
expression for the modified distribution for electrons
\begin{eqnarray}\label{tildefe}
&&\tilde{f}_{0e}({\bf r})=\int
d^3q\,q_x\,f_{0e}(\textbf{q})\,e^{-i\textbf{qr}}
\end{eqnarray}
\noindent with
\begin{equation}
f_{0e}({\bf q})=\sqrt{2\pi}\left(\frac{\hbar^2}{2\pi k_B{\mathcal
T}}\right)^{2}\exp\left(-\frac{\hbar^2q^2}{2m_e{ k_B {\mathcal
T}}}\right),
\end{equation}
where $\mathcal T$ is the temperature and $k_B$ is the Boltzmann
constant. The integral (\ref{tildefe}) can be evaluated
analytically yielding
\begin{eqnarray}\label{edistribution}
&&\tilde{f}_{0e}(\textbf{r})=\tilde{f_{0e}}(\rho,z_e,z_h,\phi)=\sqrt{\frac{\pi}{2}}\frac{r}{\lambda_{\hbox{\tiny
th e}}}\nonumber\\
&&\times
[\Phi_1(\phi)+\Phi_{-1}(\phi)]\exp\left(-\frac{r^2+(z_e-z_h)^2}{2}\frac{m_e
k_B{\mathcal T}}{\hbar^2}\right)\nonumber\\
&&=\tilde{f}_{0e}^\perp(z_e,z_h)\tilde{f}^\parl_{0e}(r,\phi),\nonumber\\
&&\tilde{f}_{0e}^\perp(z_e,z_h)=\exp\left(-\frac{(z_e-z_h)^2}{2}\frac{m_e
k_B{\mathcal T}}{\hbar^2}\right),\\
&&\tilde{f}^\parl_{0e}(r,\phi)=\sqrt{\frac{\pi}{2}}\frac{r}{\lambda_{\hbox{\tiny
th e}}}\exp\left(-\frac{r^2}{2}\frac{m_e k_B{\mathcal
T}}{\hbar^2}\right)[\Phi_1(\phi)+\Phi_{-1}(\phi)],\nonumber\\
 &&r=\sqrt{x^2+y^2},\nonumber
\end{eqnarray}
where
\begin{eqnarray}
&&\Phi_m(\phi)=\frac{e^{im\phi}}{\sqrt{2\pi}},
\end{eqnarray}
and \begin{eqnarray}\label{thermal1}
&&\lambda_{\hbox{\tiny th
e}}=\left(\frac{\hbar^2}{m_e k_B{\mathcal
T}}\right)^{1/2}=\sqrt{\frac{2\mu}{m_e}}\sqrt{\frac{R^*}{k_B{\mathcal
T}}}a^*,\end{eqnarray} is the so-called thermal length (here for
electrons).
Similarly, for the hole equilibrium distribution we have
\begin{eqnarray}\label{edistribution2}
&&\tilde{f}_{0h}(\textbf{r})=\tilde{f_{0h}}(\rho,z_e,z_h,\phi)=\sqrt{\frac{\pi}{2}}\frac{r}{\lambda_{\hbox{\tiny
th h}}}\nonumber\\
&&\times
[\Phi_1(\phi)+\Phi_{-1}(\phi)]\exp\left(-\frac{r^2+(z_e-z_h)^2}{2}\frac{m_h
k_B{\mathcal T}}{\hbar^2}\right)=\nonumber\\
&&=\tilde{f}_{0h}^\perp(z_e,z_h)\tilde{f}^\parl_{0h}(r,\phi),\\
&&\tilde{f}_{0h}^\perp(z_e,z_h)=\exp\left(-\frac{(z_e-z_h)^2}{2}\frac{m_h
k_B{\mathcal T}}{\hbar^2}\right),\nonumber\\
&&\tilde{f}^\parl_{0h}(r,\phi)=\sqrt{\frac{\pi}{2}}\frac{r}{\lambda_{\hbox{\tiny
th h}}}\exp\left(-\frac{r^2}{2}\frac{m_h k_B{\mathcal
T}}{\hbar^2}\right)[\Phi_1(\phi)+\Phi_{-1}(\phi)],\nonumber
\end{eqnarray}
\noindent with the hole thermal length
\begin{eqnarray*}&&\lambda_{\hbox{\tiny th
h}}=\left(\frac{\hbar^2}{m_h k_B{\mathcal
T}}\right)^{1/2}=\sqrt{\frac{2\mu}{m_h}}\sqrt{\frac{R^*}{k_B{\mathcal
T}}}a^*.\end{eqnarray*}
 The matrices $C$
and $D$ are temperature-dependent, so they also can  be used as an
additional contribution for interpretation of temperature
variations of excitonic optical spectra. However, the temperature dependence of relaxation constants $\Gamma_n$ remains a dominant mechanism influencing the spectra. Further, we will assume
that our medium is excited homogeneously in ${\bf X}$ space. For
$p$ excitons the matrices $C$ and $D$ relax to their values at
$r=r_0$. In Cu$_2$O, the dipole
density can be approximated by $\textbf{M}(\textbf{r})\propto
\textbf{r}\delta(r-r_0)$\cite{StB87}, which leads to the
following expressions for the matrices $C,~D$
\begin{eqnarray}
&&C({\bf r})=-\frac{i}{\hbar}\left[\tau J_C({\bf r})-\tau
J_C(r_0)+T_1f_{0e}({\bf r})J_C(r_0)\right],\nonumber\\
&&\\ & &D({\bf r})=-\frac{i}{\hbar}\left[\tau J_{V}({\bf r})-\tau
J_{V}(r_0)+T_{1}f_{0hH}({\bf r})J_{V}(r_0)\right].\nonumber
\end{eqnarray}
With the above expression the equation for the third order coherent amplitude
$Y^{(3)}_{12}$ takes the form
\begin{eqnarray}\label{jotplus}
& &\hbar\left(\omega+\frac{\rm
i}{T_{2}}\right)Y^{(3)}_{12-}-H_{QW}
Y^{(3)}_{12-}\nonumber\\
&&=M_{0}\rho_0({E}_1C_{12}+{E}_2D_{21})={E}({\bf
R},t)\tilde{J}_{-},\nonumber
\\
& &\hbar\left(-\omega+\frac{\rm
i}{T_{2}}\right)Y^{(3)}_{12+}-H_{QW}
Y^{(3)}_{12+}\nonumber\\
&&=M_{0}\rho_0({E}^*_1C_{12}+{E}^*_2D_{21})={E}^*({\bf
R},t)\tilde{J}_{+}.
\end{eqnarray}
\noindent To define the source terms $\tilde{J}_{\pm}$ 
we use the fact that for most semiconductors $T_1\gg \tau$.
Therefore we retain only the terms proportional to $T_1$,
obtaining
\begin{eqnarray}
&&\tilde{J}_{-}=-\frac{i}{\hbar}T_1M_{0}\rho_0\biggl\{J_C(r_0)\tilde{f}_{0e}({\bf
r})\nonumber\\ &&+J_{V}(r_0)\tilde{f}_{0h}({\bf r})\biggr\},\nonumber\\&&\\
 &
&\tilde{J}_{+}=-\frac{ i}{\hbar}M_{0}\rho_0T_1\biggl\{
J_C(r_0)\tilde{f}_{0e}({\bf r}+J_{V}(r_0)\tilde{f}_{0h}({\bf
r})\biggr\}.\nonumber
\end{eqnarray}
\noindent From $Y^{(3)}$ one finds the third order polarization
according to
\begin{eqnarray}
&&P^{(3)}({\bf R})=2\int {\rm d}^3{r} \hbox{Re}\, M({\bf r})
Y^{(3)}({\bf R},{\bf r})\nonumber
\\&&=\int {\rm d}^3{r}\,M({\bf r})\left(Y^{(3)}_{12-}+Y^{(3*)}_{12+}\right).
\end{eqnarray}
\noindent As in the case of linear amplitudes $Y^{(1)}$, we expand
the nonlinear amplitudes in terms of the eigenfunctions
$\Psi_{\ell mN_eN_h}({\bf r})$.

The next application of the amplitude $Y_{12}^{(1)}$ is related to the
iteration process. Inserting $Y_{12}^{(1)}$ in the source terms on the
r.h.s. of Eqs. (\ref{conduction1},\ref{valence1}) and using
appropriate expressions for the irreversible terms, one obtains
the matrices $C^{(2)}, D^{(2)}$, where the superscript indicates
the order with respect to the electric field strength $E$.
Substituting the matrices into the saturating terms on the r.h.s.
of Eq. (\ref{interband1}) one obtains the equation for the
nonlinear amplitude $Y^{(3)}$ which, with respect to Eq.
(\ref{polarization1}), defines the nonlinear susceptibility
$\chi^{(3)}$. We obtain the following expression
\begin{eqnarray}\label{chi3}
&&\chi^{(3)}(\omega)=-\left(\frac{a^*}{L}\right)\chi_0^{(3)}\\
&&\times\sum\limits_{j,\ell,N}\frac{{\mit\Gamma}_j{\mathcal
F}_{j\ell N}E_{T\ell
N}}{[(E_{TjN}-\hbar\omega)^2+{\mit\Gamma_j}^2][E_{T\ell
N}^2-(E+i{\mit\Gamma}_{\ell})^2]},\nonumber
\end{eqnarray}
where $\ell=0,1,2...$ The nonlinear oscillator strengths can be written as
\begin{eqnarray}\label{non_osc}
&&{\mathcal F}_{j\ell
N}=\frac{M(-j,3,4\kappa_{j}\rho_0)\,M\left(-\ell,3,4\kappa_{\ell
}\rho_0\right)}{(1+2\kappa_{\ell
}\rho_0)^4(1+2\kappa_{j}\rho_0)^4}\nonumber\\
&&\times
\,F\left(-j,4;3;\frac{4\kappa_{j}\rho_0}{1+2\kappa_{j}\rho_0}\right)\,F\left(-\ell,4;3;\frac{4\kappa_{\ell
1}\rho_0}{1+2\kappa_{\ell}\rho_0}\right)\nonumber\\
&&\times
\frac{(j+1)(j+2)}{(2j+3)^5}\,\frac{(\ell+1)(\ell+2)}{(2\ell+3)^2}\left[{\mathcal
A}\,\left(\frac{2}{2\ell+3}\right)^\beta\,V_{NN}^{(e)}\right.\nonumber\\
&&\left. +{\mathcal
B}\,\left(\frac{2}{2\ell+3}\right)^\gamma\,V_{NN}^{(h)}\right],\nonumber\\
&&\kappa_{j}=\frac{1}{2j+3},\quad\kappa_{\ell
}=\frac{1}{2\ell+3},
\end{eqnarray}
where 
\begin{equation}
\chi_0^{(3)}= \epsilon_0(\epsilon_b\Delta_{LT})^2a^{*3}e^{-4\rho_0}\left(\frac{1}{{\mit\Gamma}_{1}}\right)
\end{equation}
and the derivation of constants $\mathcal A$, $\mathcal B$ is presented in Appendix \ref{App1}. The potentials $V_{N_eN_h}$ are given by 
\begin{eqnarray}\label{eq:potentials}
&&V_{N_eN_h}^{(e)}=\int\limits_{-1}^{1}
dx\,\int\limits_{-1}^{1}\,dy
\cos\left[\frac{(2N_e-1)\pi}{2}x\right]\cos\left[\frac{(2N_h-1)\pi}{2}y\right]\nonumber\\
&&\times\exp\left[-\left(\frac{L^2}{8a^{*2}{\tilde{\lambda}_{\hbox{\tiny
th e}}^2}}\right)(x-y)^2\right]\nonumber\\
&&V_{N_eN_h}^{(h)}=\int\limits_{-1}^{1}
dx\,\int\limits_{-1}^{1}\,dy
\cos\left[\frac{(2N_e-1)\pi}{2}x\right]\cos\left[\frac{(2N_h-1)\pi}{2}y\right]\nonumber\\
&&\times\exp\left[-\left(\frac{L^2}{8a^{*2}{\tilde{\lambda}_{\hbox{\tiny
th h}}^2}}\right)(x-y)^2\right],
\end{eqnarray}
with the  thermal lengths $\tilde{\lambda}_{\hbox{\tiny th e,h}}$ defined above in Eq. (\ref{thermal1}).

More detailed calculations of $\chi^{(3)}$ are presented in Appendix \ref{App2} and the table of material parameters is included in Appendix \ref{App3}. The long-wave approximation we have used in above calculations limits our results, which are appropriate for quantum wells thicknesses $L$ significantly bellow $1 \mu m$.

\section{Rydberg blockade}\label{SectBlock}
One of the important characteristics of the theoretical approach described above is the fact that it is derived under the assumption of a relatively low power level, when the medium is not saturated with excitons. Thus, the so-called Rydberg blockade \cite{Kazimierczuk2014} is not inherently present in the calculations and its effects have to be taken into account in a separate step. This has been done in \cite{Ziemkiewicz2019,ThomasArxiv2022} and the description outlined below is an extension of the approaches presented in the cited works.

\begin{figure}[ht!]
\includegraphics[width=.7\linewidth]{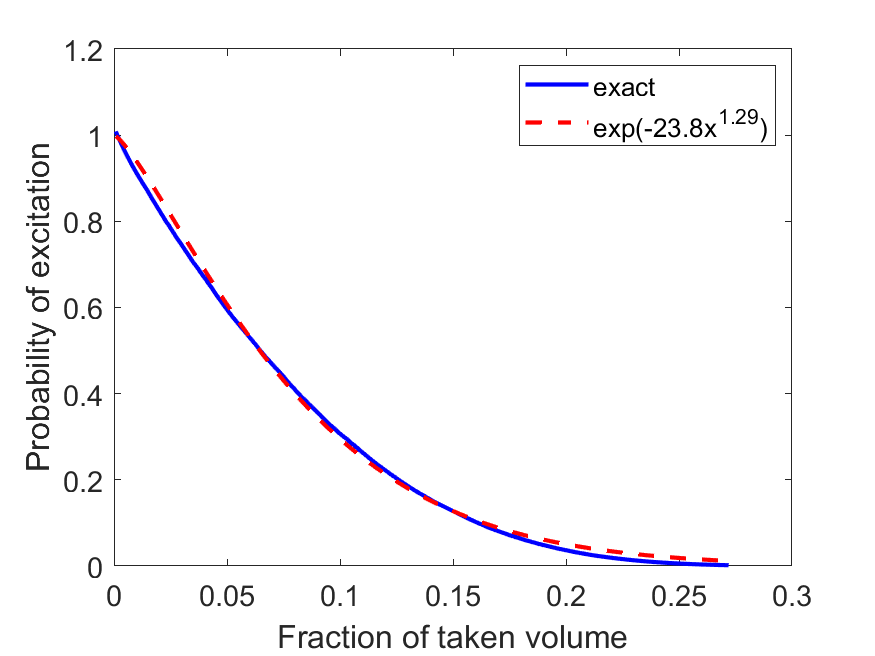}
\caption{The probability of excitation as a function of the volume occupied by Rydberg blockade.}\label{fig:spherepack}
\end{figure}

For an exciton with principal quantum number $(j+1)$, the blockade volume is given by \cite{Kazimierczuk2014}
\begin{equation}
V_B = 3 \cdot 10^{-7} (j+1)^7~\mu m^3.
\end{equation} 
Similarly to the recent experiments \cite{ThomasArxiv2022}, we assume that the laser beam illuminating the sample has a circular beam spot of area $S$ of 0.1 mm$^2$ and  the sample length is $L$; the volume , where the light can be absorbed and an exciton created is $V=LS$. Within this volume, a new exciton can be formed only when its location is outside of the blockade volume of existing excitons. Thus, assuming that the blockade volume is spherical, the upper limit of exciton density is the perfect sphere packing, where approximately $74\%$ of the volume is occupied, e. g. for the number of excitons $N_e$, $N_e\frac{V_B}{V} \approx 0.74$. However, the positions of the excitons formed within the laser beam are random and thus highly unlikely to form a perfect sphere packing. To estimate the practical upper limit of exciton density imposed by Rydberg blockade, a Monte Carlo simulation has been performed; within given volume $V$, excitons with their associated blockade volumes are added at random positions and the number of attempts to place an exciton in a free space (not occupied by blockade volume) is counted. Then, the probability of excitation (inverse of the number of attempts) is calculated. The results are shown on the Fig. \ref{fig:spherepack}.
One can see that the system is effectively saturated when the fraction of occupied volume approaches 0.2. An exponential function can be fitted to the data (dashed line), providing a simple model of saturation; the probability of excitation is
\begin{equation}\label{satmodel}
p_e \approx \exp\left(-23.8\left[\frac{N_eV_B}{V}\right]^{1.29}\right).
\end{equation} 
When calculating the susceptibility from Eq. (\ref{chilin1}) and Eq. (\ref{chi3}), one has to multiply the oscillator strengths $F$ by the above probability. This is a similar approach to that one used in \cite{ThomasArxiv2022} and \cite{Ziemkiewicz2019}, where also an exponential function $\exp(-AN_eV_B/V)$ with one fitted constant $A$ was used.

Finally, to calculate the number of excitons (and thus the blocked volume), one can consider the power to sustain a single exciton
\begin{equation}
P_1 = \frac{E_j}{\tau_j}
\end{equation}
where $E_j$ and $\tau_j$ are the energy and lifetime of excitonic state. The number of excitons $N_e$ is
\begin{equation}
N_e = \frac{P_A}{P_1}
\end{equation}
where $P_A$ is the absorbed laser power; for a sufficiently thick sample, it is equal to the total laser power.

\begin{figure}[ht!]
\centering
a)\includegraphics[width=.85\linewidth]{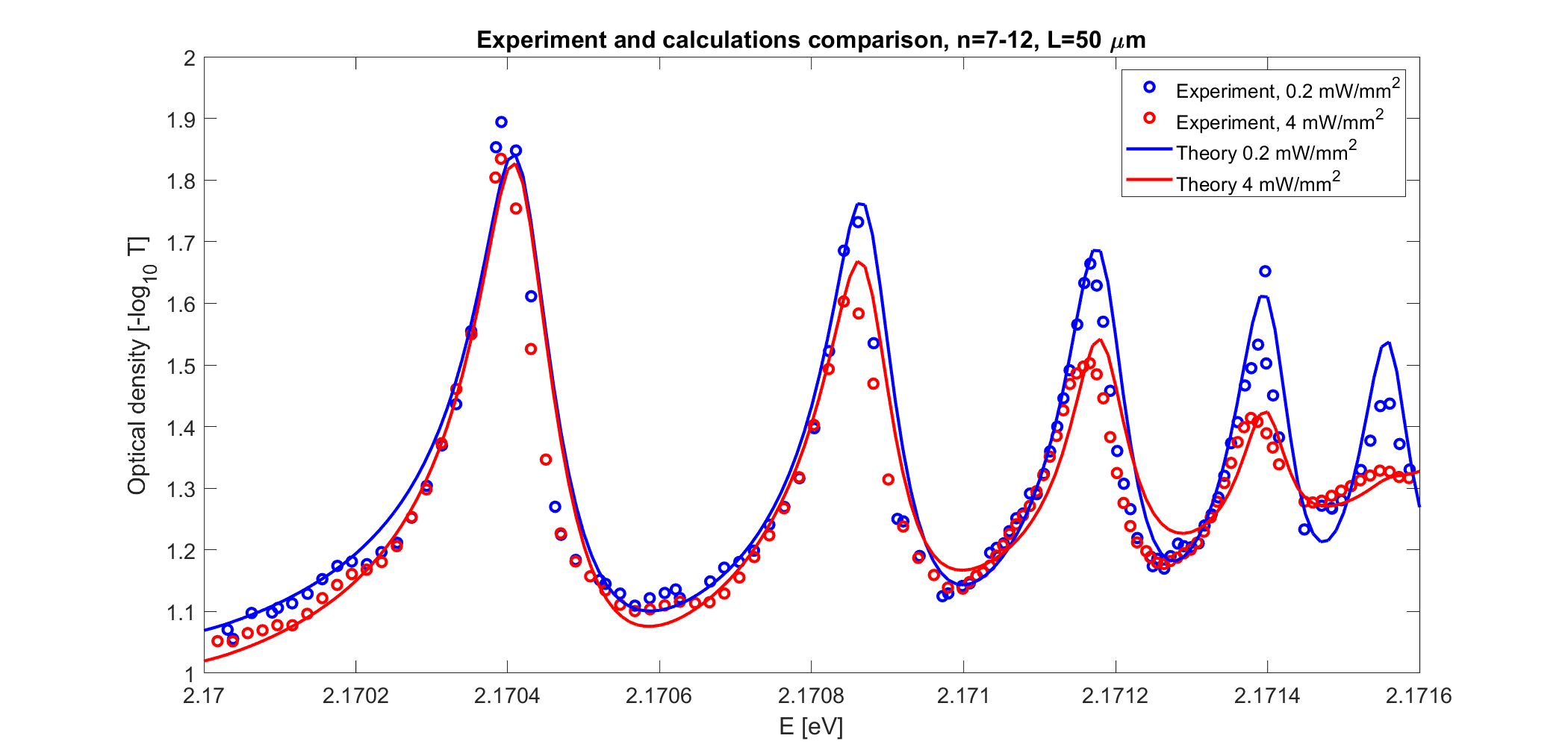}
b)\includegraphics[width=.85\linewidth]{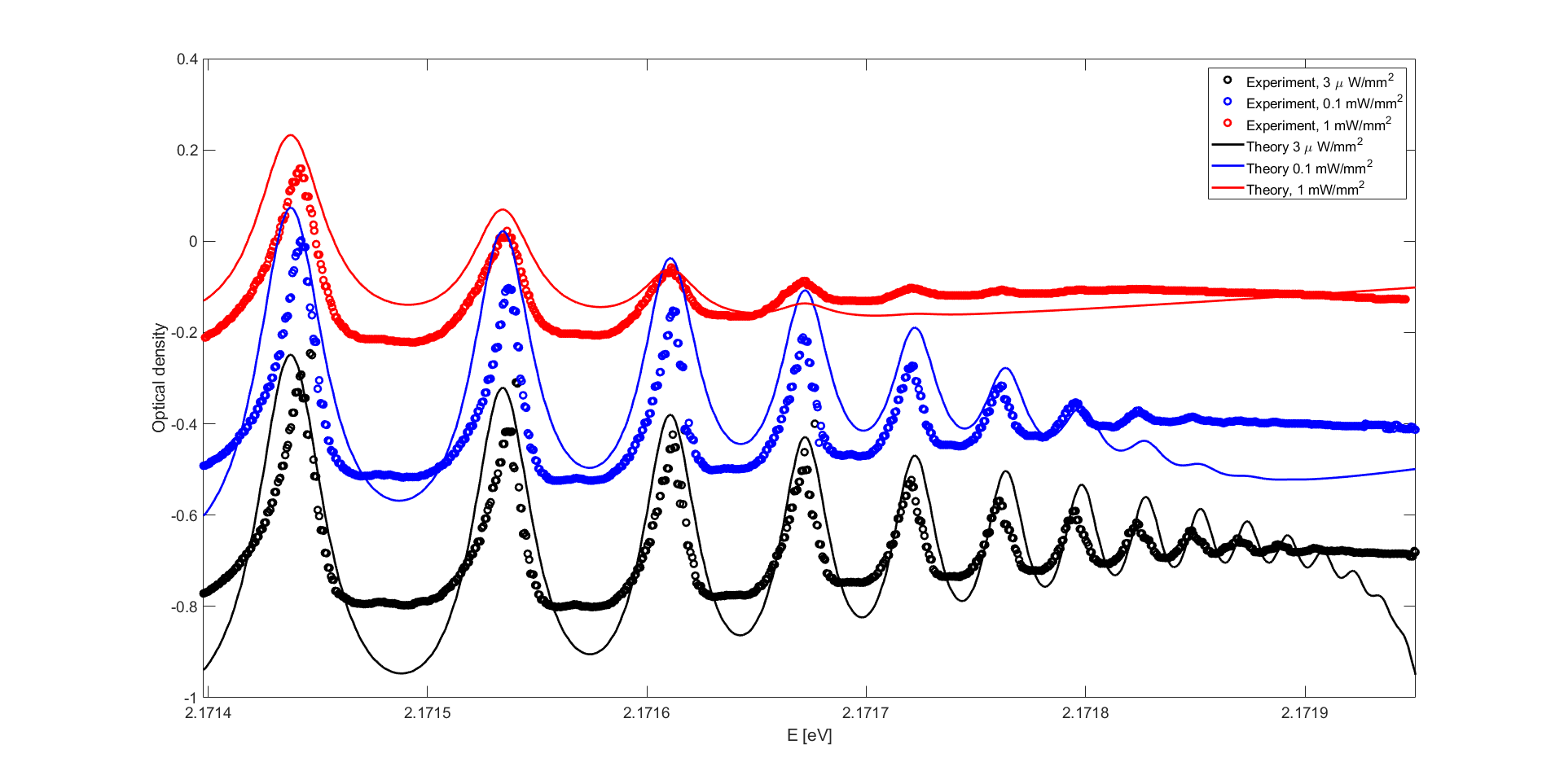}
\caption{a) Calculated optical density, compared to experimental results in bulk crystal \cite{ThomasArxiv2022}; b) Absorption spectrum compared to the data from \cite{Kazimierczuk2014}.}
\label{porThomas}
\end{figure}

As a first verification of the presented theoretical description, one can examine the results obtained in the asymptotic limit of a very large thickness, e.g. a bulk crystal. The results of such a comparison are presented on the Fig.\ref{porThomas}. Specifically, Fig. \ref{porThomas} a) depicts the calculated optical density spectrum in the region of $j=6-11$ excitonic states, for two illumination powers. One can notice a quick decrease of absorption in the high power regime, approximately proportional to blockade volume $\sim j^7$. This is the so-called optical bleaching \cite{Kazimierczuk2014}. The same result can be seen on the Fig. \ref{porThomas} b), where calculations are compared to the experimental data from \cite{Kazimierczuk2014}. A very good agreement obtained in a wide range of powers and across multiple excitonic states indicates that the saturation model in Eq. (\ref{satmodel}) is sufficiently precise.
\section{Self-Kerr nonlinearity}
In the self Kerr effect the refractive index is
changed due to the response of the incoming field itself, in other words it consists in the change of the refractive index of the medium with a variation of the propagating light intensity.  
The third-order nonlinear susceptibility is the  basis of theoretical description of this phenomenon.  The nonlinear optical response is conveniently described in terms of a field-dependent index $n(E)$ defined as
\begin{equation}
n^2(E)=1+\chi=\epsilon_b+\chi^{(1)}+\chi^{(3)}|E|^2+... .
\end{equation}
The real part of the nonlinear  susceptibility defines the nonlinear index of refraction, which characterizes so-called Kerr media, 
$n_2=\frac{Re \chi^{(3)}}{c\epsilon_0n_0^2}$,
with $n_0^2=1+\chi^{(1)}$.
\\\noindent
The self-Kerr interaction is an optical nonlinearity that produces a phase shift proportional to
the square of the field intensity (or a number of photons in the field). 
In the Kerr medium the phase of an electromagnetic wave propagating at the distance $L$ increases and the increment in phase due to intensity-dependent term is proportional to the distance and to  the square of the electric field strength, which is called self-phase modulation.
 The phase shift is calculated from
\begin{equation}\label{deltafi}
\Delta\Phi=\frac{\omega L}{c}[n(|E|^2)-n(0)]
\end{equation}
The considerable nonlinear susceptibility of Rydberg excitonic system, further amplified in a thin quantum well, is expected to cause a noticeable phase shift even for small $L \sim 100$ nm. The confinement states, even when not directly visible, still contribute to the total height of the excitonic line, increasing $\chi^{(3)}$ and phase shift.
\section{Results}
Due to the limited amount of experimental data regarding nonlinear properties of Cu$_2$O quantum wells, as a first step we verify our calculations with a comparison to a bulk medium. It should be stressed that while the calculated spectra approach the bulk ones as $L \rightarrow \infty$, the presented method is derived under the assumptions of strong confinement and long-wave approximation, so it yields fully correct results only for well thickness significantly below $1\mu$m. In contrast to \cite{Konzelmann2019}, where weak confinement regime is studied, the confinement affects the relative motion of the electron-hole pair.

\begin{figure}[ht!]
\centering
a)\includegraphics[width=.8\linewidth]{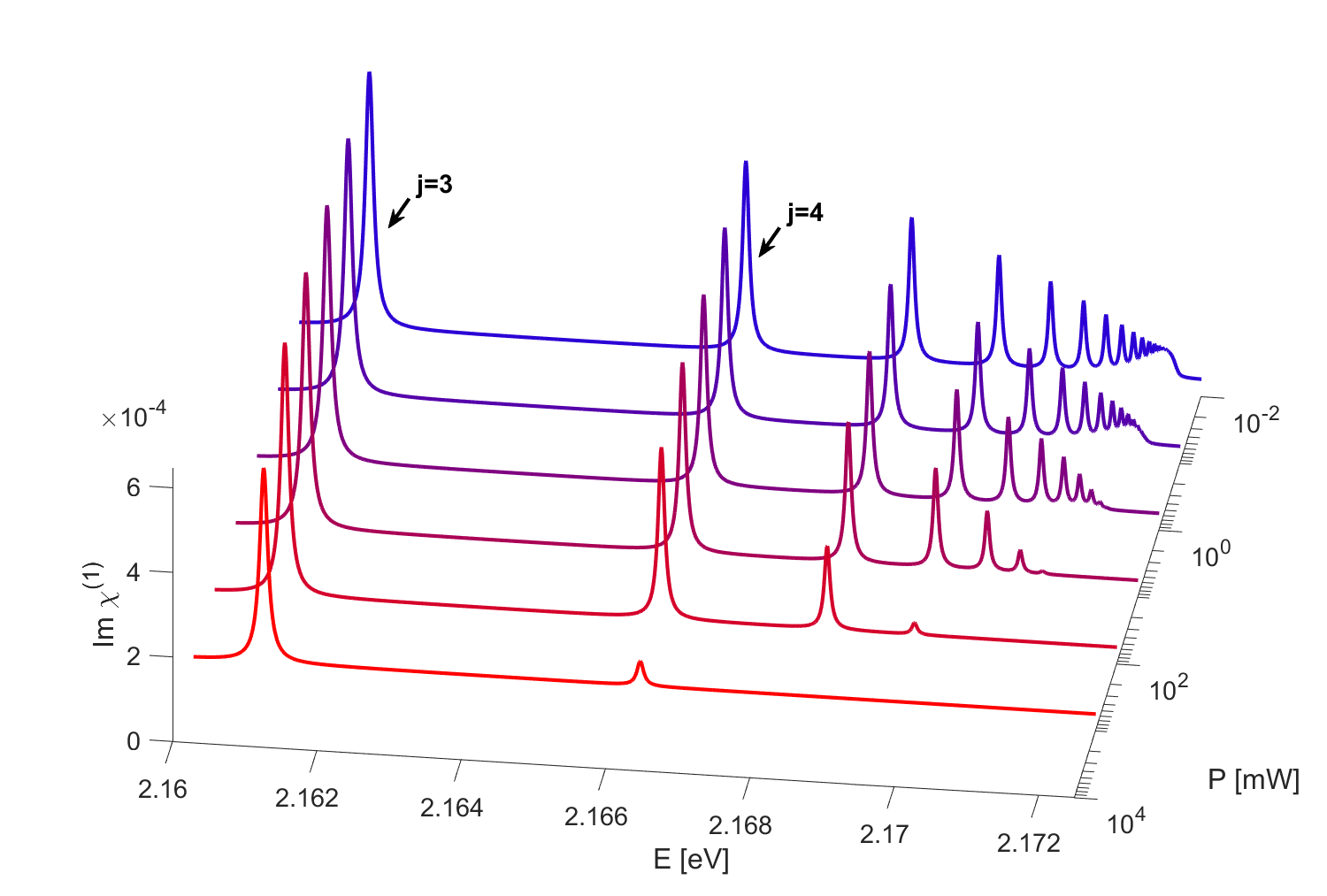}
b)\includegraphics[width=.8\linewidth]{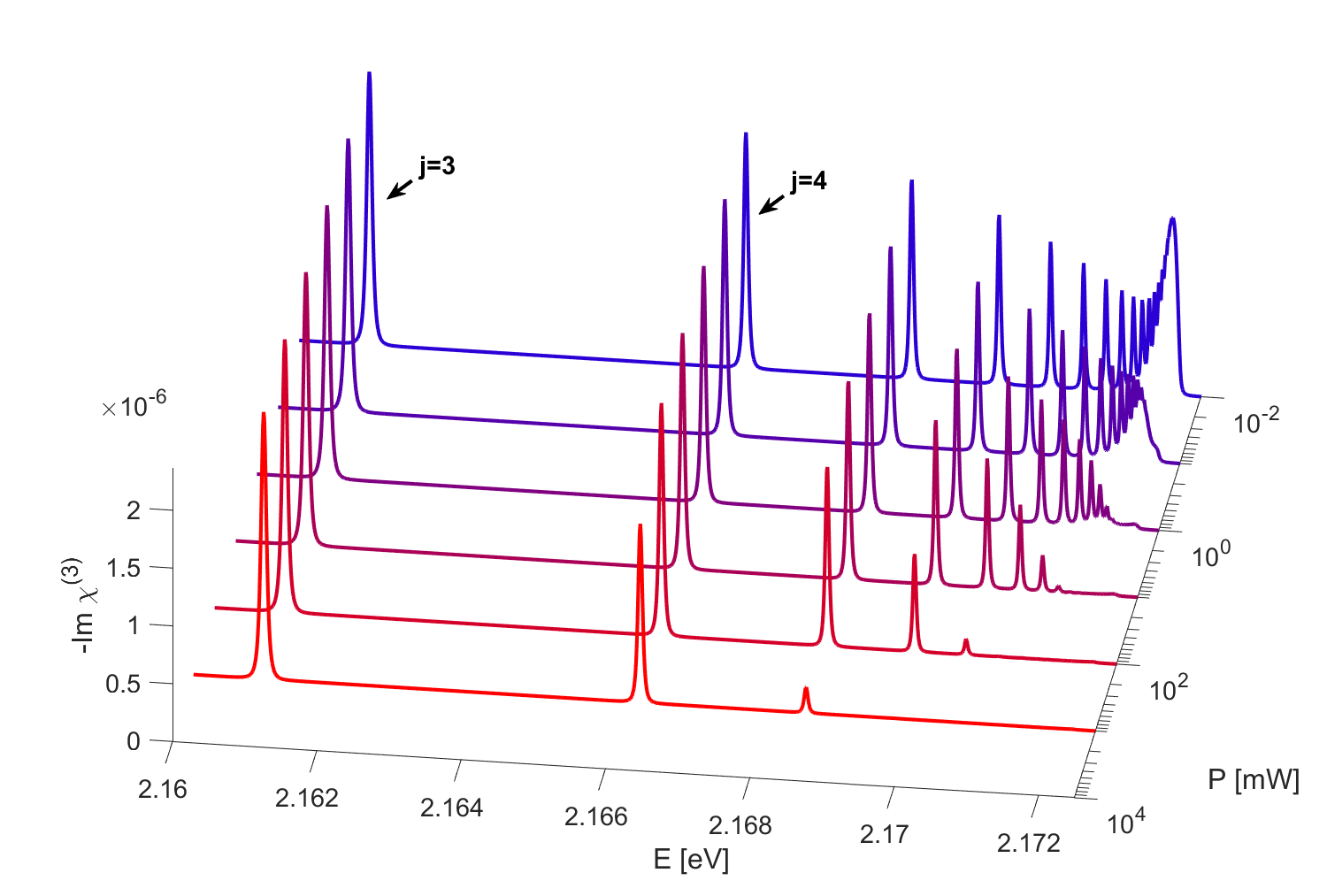}
\caption{Imaginary parts of linear $\chi^{(1)}$ and nonlinear $\chi^{(3)}$ susceptibility in a bulk crystal ($L=50~\mu$m), for selected laser powers.}\label{fig:powbulk}
\end{figure}

As mentioned above, one can make a rough comparison with experimental results in bulk medium by assuming a large value of $L$, skipping the wide quantum well regime at moderate $L \sim 1$ $\mu$m. The linear and nonlinear parts of susceptibility have been calculated from Eqs. (\ref{chilin1}) and (\ref{chi3}), in a wide range of laser powers and for a thick crystal $L=50$ $\mu$m. The results are shown on the Fig. \ref{fig:powbulk}. 
The calculated spectra are in the range from $j=3$ exciton resonance (2.161 meV) to the band gap (2.172 meV). The effect of Rydberg blockade is included in calculations by multiplying the obtained susceptibility by the factor $p_e$, Eq.(\ref{satmodel}); as the power increases, the density of excitons reaches saturation and $p_e \rightarrow 0$. In such a way one is able to control  whether one is still in the regime, in which additional effects due to Rydberg blockade preventing the transmission are absent, and do not influence the excitons-light interaction. 
 One can see that the overall amplitude of the linear susceptibility changes  in the order of $10^{-4}$ and the nonlinear part is approximately 3 orders of magnitude lower. As expected, the number of observed resonances is strongly dependent on the power $P=\frac{1}{2}cS_b\epsilon|E|^2$, where $S_b=0.1$ mm$^2$ is the beam area; for $P=1$ W, the bleaching is considerable even for $j=3$. The results are consistent with our previous calculations in a bulk medium \cite{ThomasArxiv2022,Raczynska2019}, as well as experimental observations \cite{Heckotter2018} and indicate that the model in Eq. (\ref{satmodel}) is correct.

\begin{figure}[ht!]
\centering
a)\includegraphics[width=.8\linewidth]{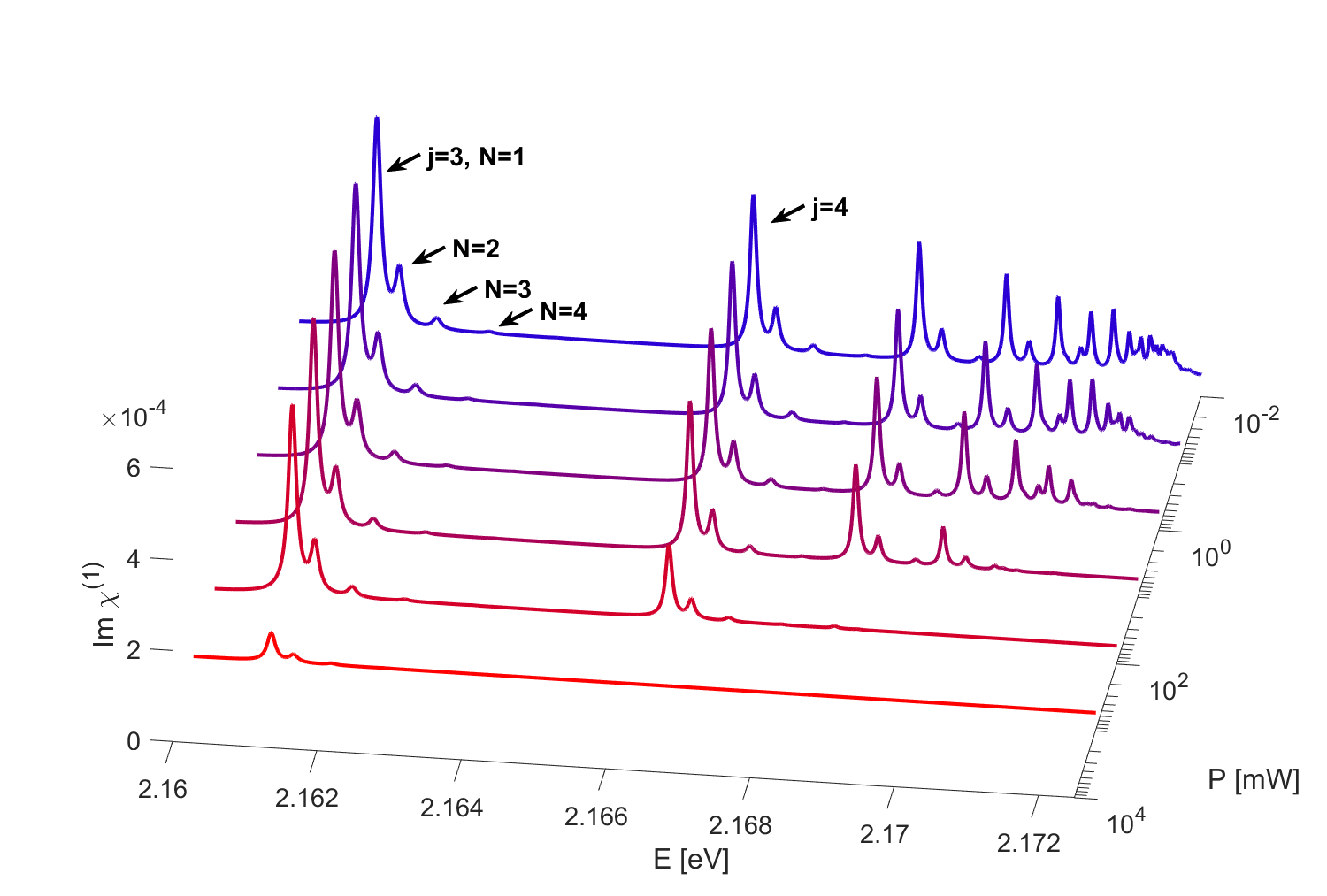}
b)\includegraphics[width=.8\linewidth]{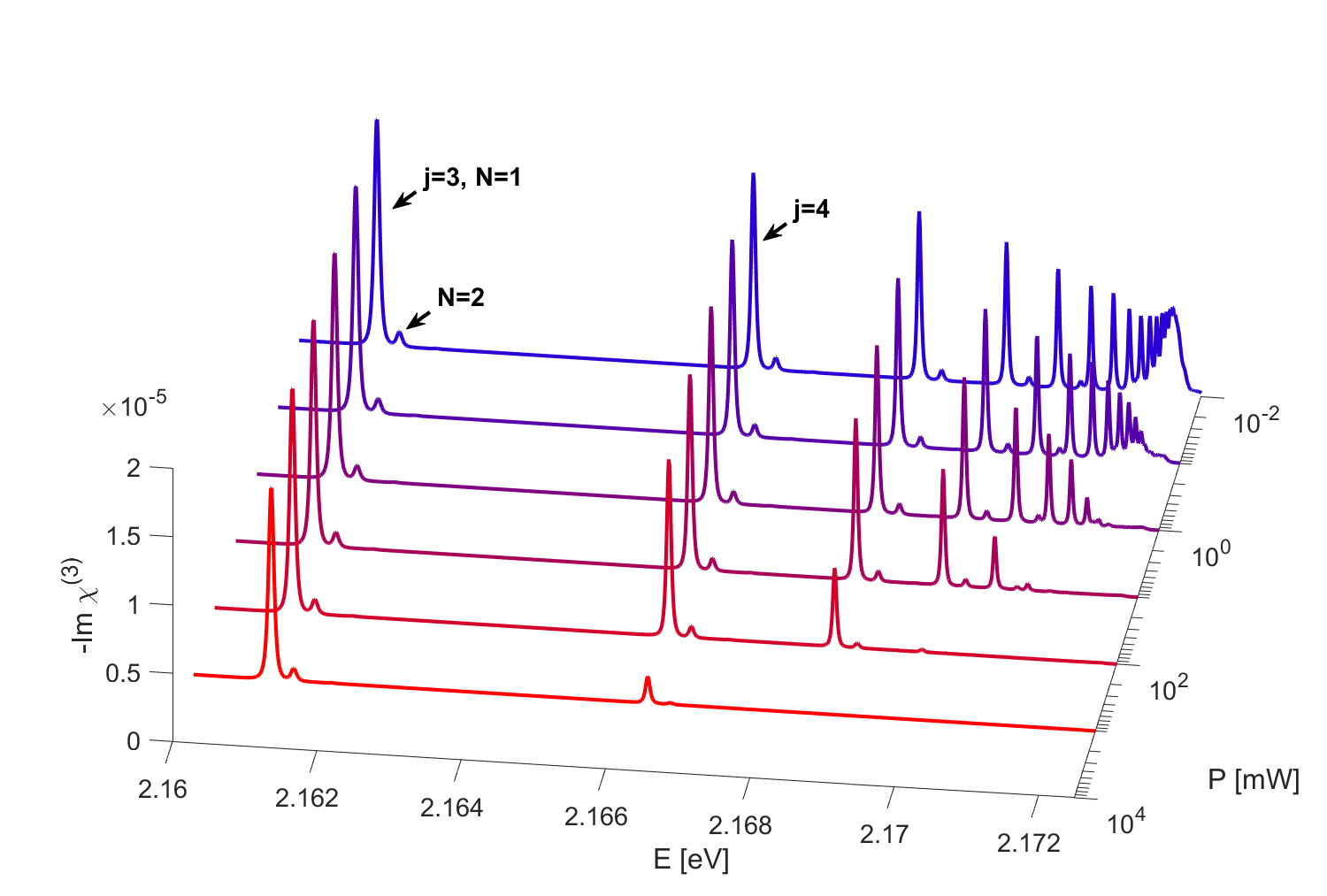}
\caption{Imaginary parts of linear $\chi^{(1)}$ and nonlinear $\chi^{(3)}$ susceptibility in $L=100$ nm quantum well, for selected laser powers.}\label{fig:powwell}
\end{figure}

As the next step, let us consider a thin $L=100$ nm quantum well. In such a system, one can expect that the absorption spectrum will contain multiple confinement states corresponding to the quantum number $N=1,2,3...$. This is the case shown in Fig. \ref{fig:powwell}. Although there is no strict upper limit on the confinement state number $N_{max}$, in practice only a few lowest confinement states are observable and thus in calculations one can assume $N_{max}=10$. The linear part of susceptibility is consistent with the results presented in \cite{Czajkowski2020} for the case of a quantum well. Specifically, one can see a series of secondary peaks originating from every excitonic line, which shift towards high energy as $L$ becomes very small. It should be stressed that these lines corresponding to confinement states are only detectable in the case of a very thin quantum well; in the micrometer-sized nanoparticles, the energy spacing between these lines is small enough that they completely overlap, forming a single, broadened excitonic line \cite{Orfanakis2021}. Moreover, in this size range, one cannot observe oscillations of the absorption coefficient caused by the spatial matching between the center-of-mass exciton motion and light waves \cite{Takahata}. Naturally, a very small absorption of a thin sample makes a direct observation of confinement states challenging. Moreover, just like in the case of large quantum dots \cite{Orfanakis2021}, the oscillator strength of excitonic states decreases faster than $j^{-3}$; this is also consistent with the observations in \cite{Konzelmann2019}. This effect, in addition to the broadening and chaotic ,,background'' formed by multiple confinement lines, puts an upper limit on the maximum principal number of the observable state.

The nonlinear susceptibility shown in the Fig. \ref{fig:powwell} b) is apparently similar to the bulk case in the Fig. \ref{fig:powbulk}.
The influence of the confinement on the nonlinear part $\chi^{(3)}$ is complex. One can see from  Eq. (\ref{eq:potentials}) that oscillating terms of $V_{N_e,N_h}$    interplay with slowly varying  factors $\lambda_{th e,h}$, describing plasma effects, resulting in an absorption attenuation.
Namely, the confinement lines are much less pronounced and only $N=1$ line is readily visible. This effect follows from Eq. (\ref{non_osc}).  

For low-dimensional systems the nonlinear optical effects depend strongly on the shape of the confinement potentials. For the above-used no-escape boundary conditions we obtained the expressions $V_{N_e,N_h}$ decaying as $N^{-1}$. The physics behind is that the rapid motion of electrons and holes in the confinement in $z$ direction, especially for states with higher $N$, hinders the creation of plasma which is responsible for the reduction of the absorption while the linear absorption does not depend on $N$. For low-dimensional systems the nonlinear optical effects depend strongly on the shape of the confinement potentials. In the considered quantum well with no-escape boundary conditions, the overall amplitude of the nonlinear part of the susceptibility is enhanced as compared to bulk system. The influence of the confinement on the nonlinear part of $\chi^{(3)}$ is more complex. One can see that  oscillating  functions $V_{{N_{e}}{N_{h}}}$, characteristic for low dimensional confined systems  interplay with relatively slowly varying exponential functions due to plasmonic terms, which results in increasing of the nonlinear absorption. 

The next Fig. \ref{fig:thickwell} shows the susceptibility spectra calculated for low a laser power and various values of thickness.
\begin{figure}[ht!]
\centering
a)\includegraphics[width=.8\linewidth]{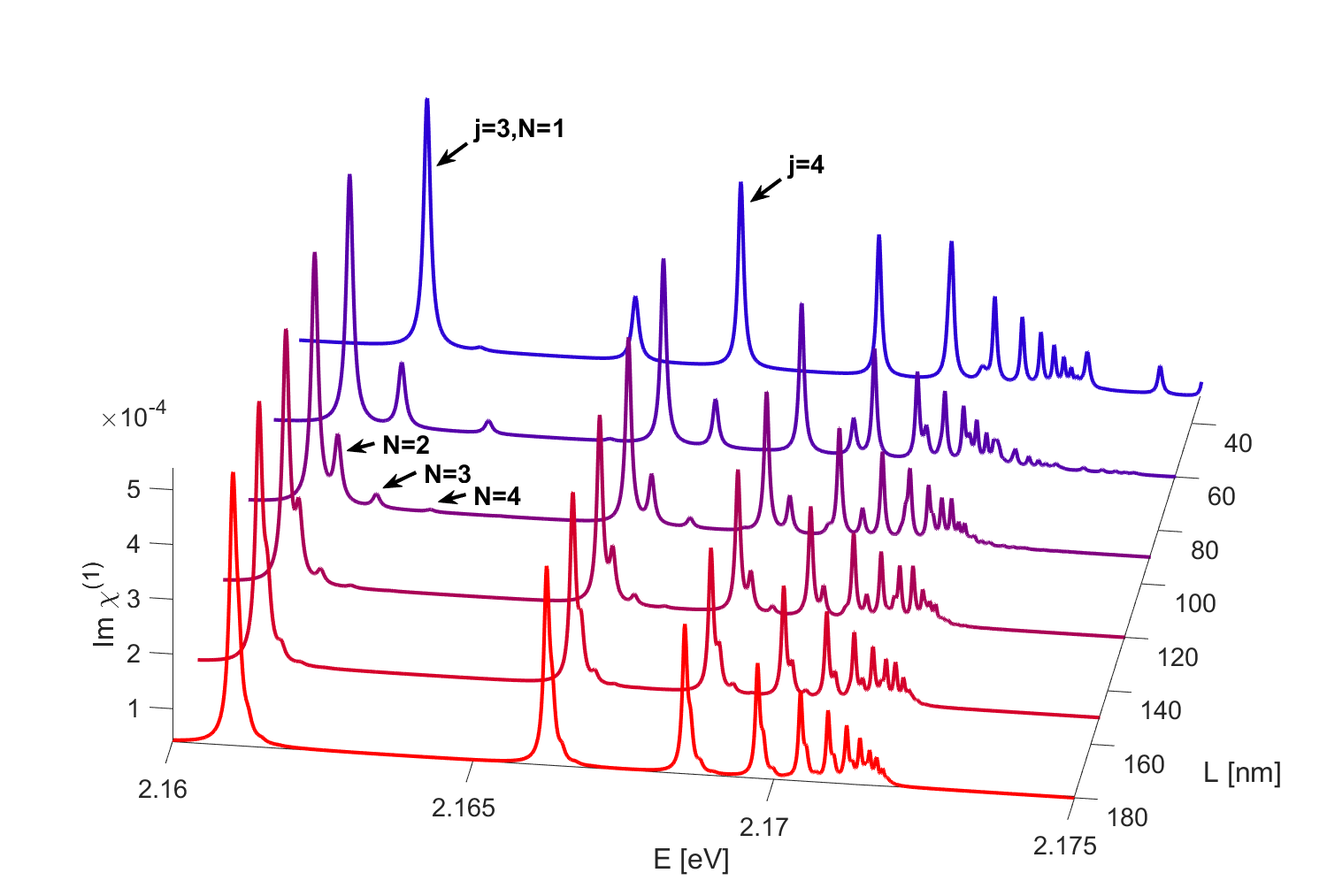}
b)\includegraphics[width=.8\linewidth]{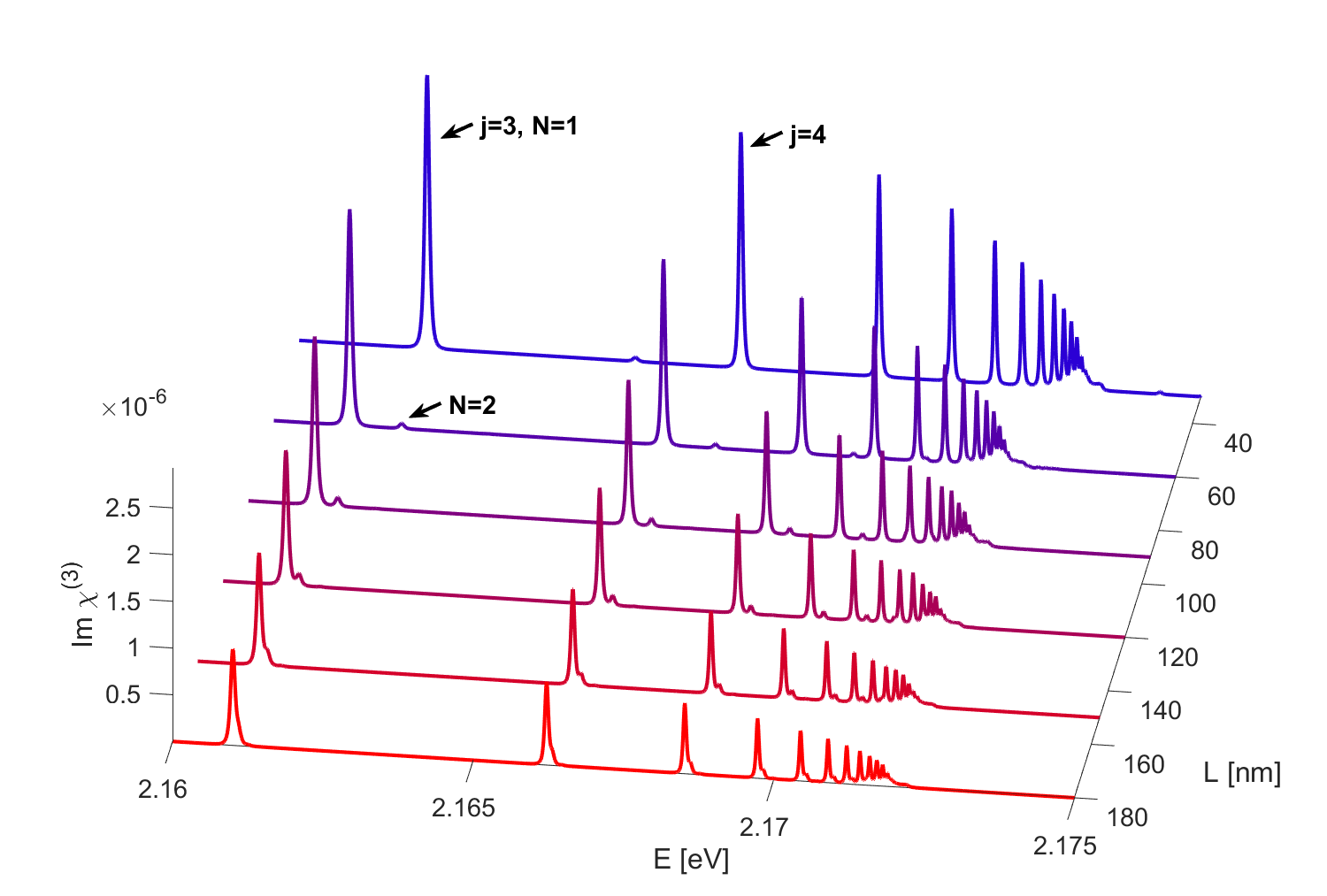}
\caption{Imaginary parts of linear $\chi^{(1)}$ and nonlinear $\chi^{(3)}$ susceptibility for selected values of quantum well thickness, for P=0.1 mW.}\label{fig:thickwell}
\end{figure}
One can see that both confinement lines and the main excitonic lines are blueshifted in the limit of small $L$; as noted in \cite{Konzelmann2019}, the confined exciton gains additional energy and this energy shift is most pronounced for $L<4a_B$, which is approximately 100 nm for $n=10$ exciton. On the other hand, it is known that excitons cannot form in quantum dots when the dot size $r<0.4 a_B$  \cite{Konzelmann2019} which indicates the lower limit of applicability of our theoretical description. As before, the lines corresponding to the confinement states are mostly invisible in the nonlinear susceptibility spectrum. In the linear part, one can see that peaks due to those states, located closely to those due to main excitonic states at $L=100$ nm , shift quickly towards higher energy for smaller $L$ because of the changing proportion between confinement energy and excitonic state energy. Due to this divergence, a considerable mixing of states occurs and also many lines can be visible in the energy region above the band gap. As mentioned before, the nonlinear part of susceptibility is enhanced in a thin quantum well; on the Fig. \ref{fig:thickwell} b) one can observe that absorption peaks become higher as $L$ decreases. The confinement of electrons and holes in a QW results in, illustratively speaking, "squeezing" of excitons, which increases the binding energy and the oscillator strength of excitons, thus leading to an enhancement of the absorption.  

\begin{figure}[ht!]
\centering
a)\includegraphics[width=.8\linewidth]{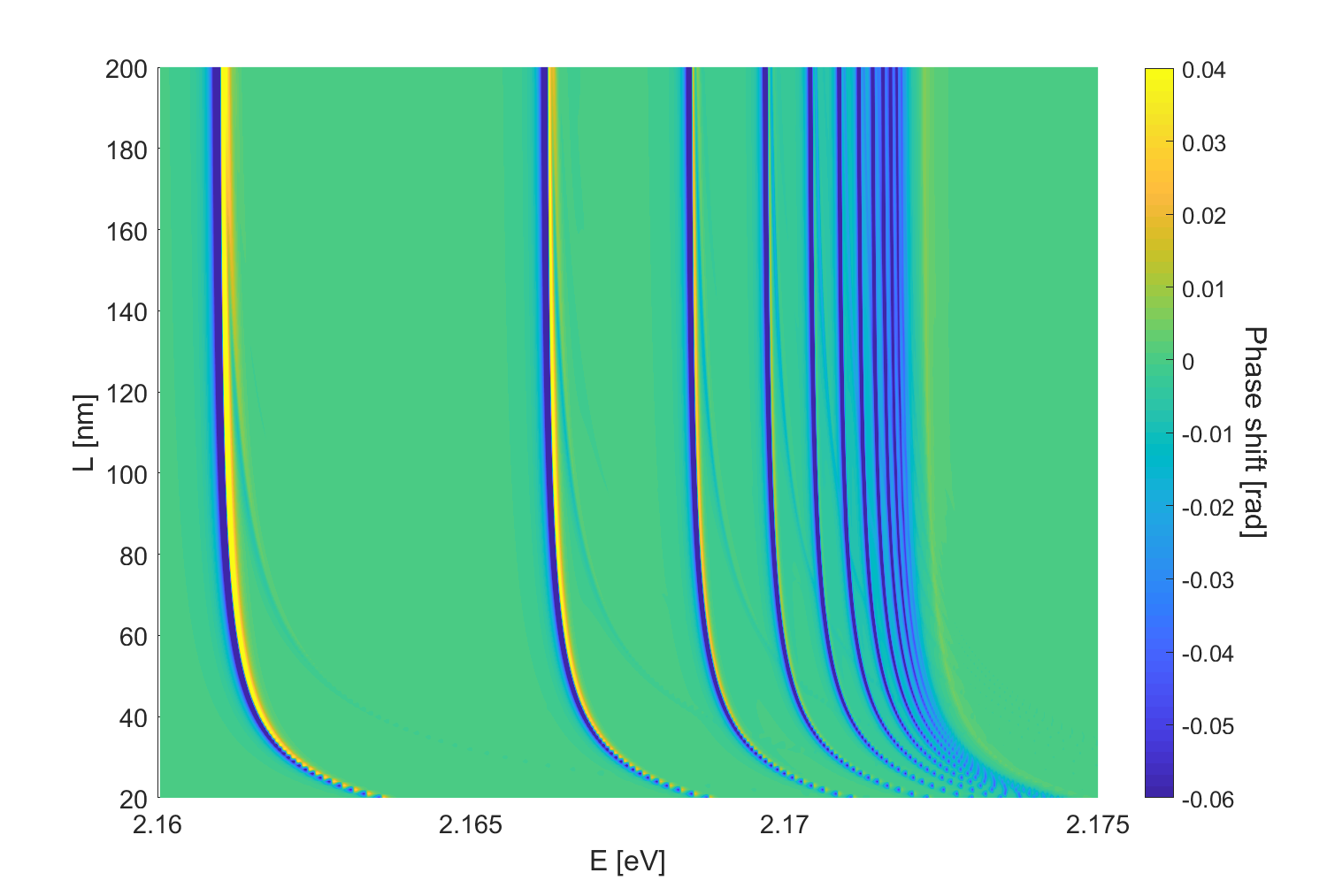}
b)\includegraphics[width=.8\linewidth]{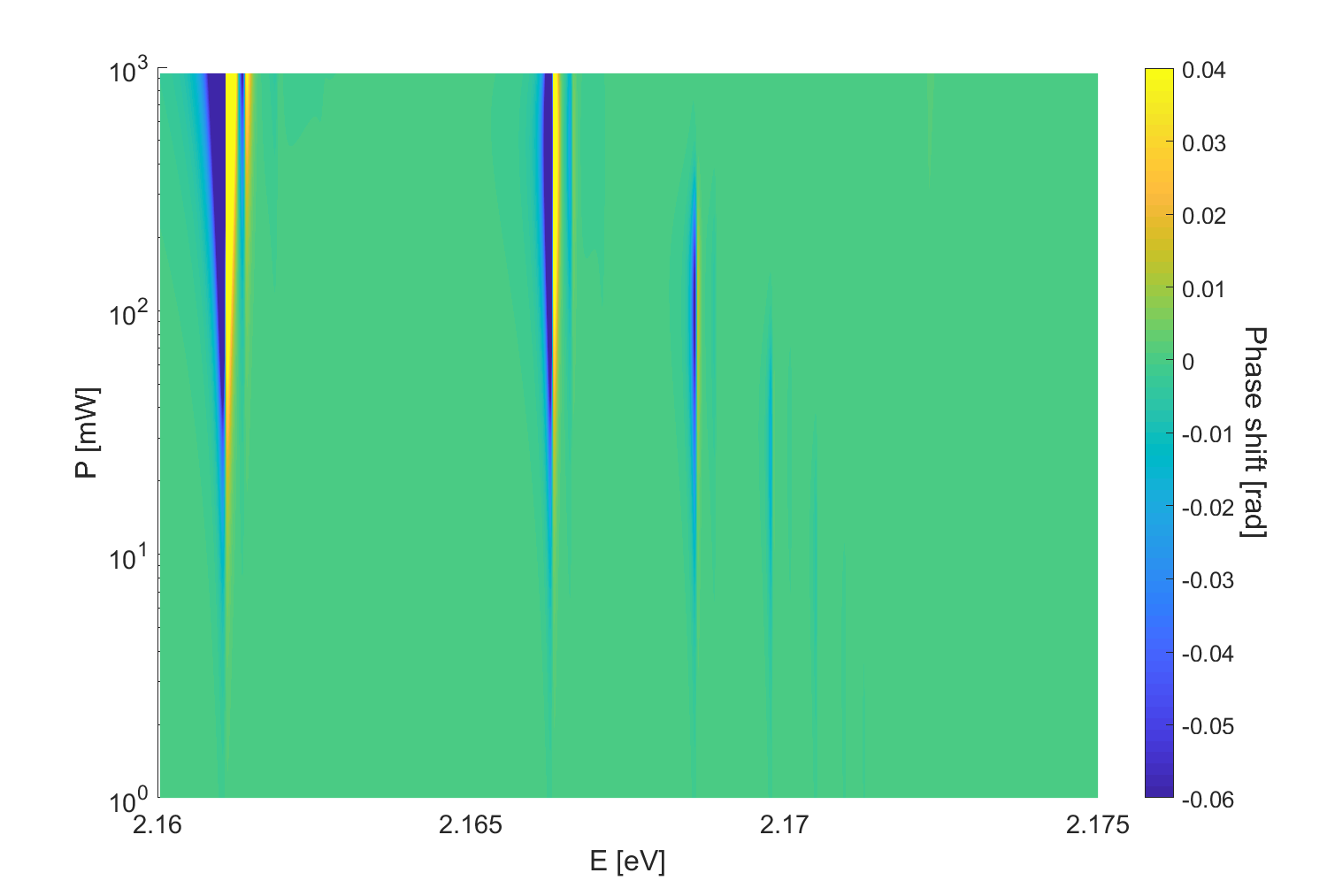}
\caption{The self-Kerr shift as a function of a) quantum well thickness and b) light power.}\label{fig:kolor1}
\end{figure}

\begin{figure}[ht!]
\centering
a)\includegraphics[width=.6\linewidth]{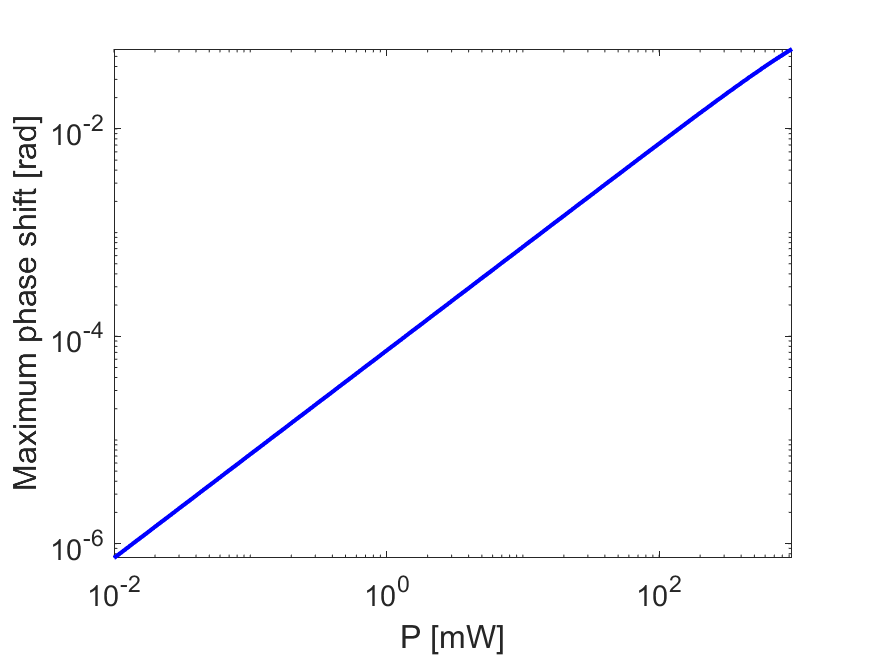}
b)\includegraphics[width=.6\linewidth]{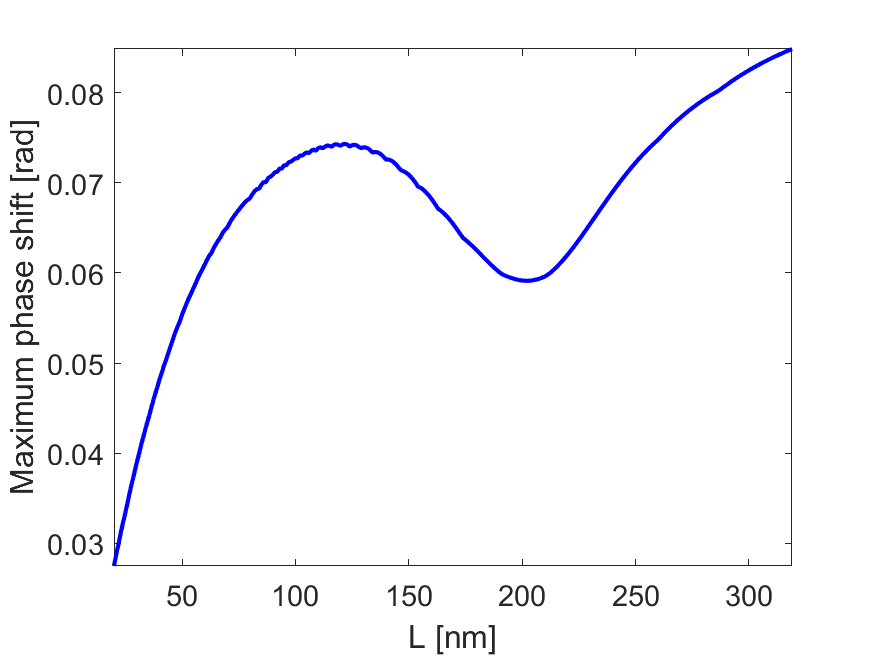}
\caption{The maximum value of self-Kerr phase shift as a function of a) light power and b) quantum well thickness.}\label{fig:ps2}
\end{figure}

Finally, we can explore the real part of susceptibility and the associated Kerr shift. Naturally, as follows from the Kramers-Kronig relations, each peak in the absorption spectrum corresponds to a region of anomalous dispersion, where $Re$ $\chi$ and thus also phase shift changes sign. This is visible in the Fig. \ref{fig:kolor1} a). As mentioned before, the confinement states are barely visible in the nonlinear part of susceptibility and thus the spectrum is dominated by lines corresponding to excitonic states $j=0,1,2...$. Again, we see a divergence towards higher energy as $L$ decreases and also a reduction of phase shift in the limit of small $L$ due to the reduced optical length $nL$ in Eq. \ref{deltafi}. Even for a relatively low thickness $L < 100$ nm, one can observe a phase shift on the order of 50 mrad. The dependence of phase shift on the laser power is shown on the Fig. \ref{fig:kolor1} b). Overall, the lower excitonic states provide a larger phase shift due to their larger oscillator strengths. The shift increases with power but is limited by optical bleaching caused by Rydberg blockade; one can see that the influence of higher states vanishes at high power. On the other hand, in the relatively lower power regime, the stronger nonlinear properties of upper states result in a considerable phase shift. 
An useful measure of the nonlinearity is the maximum phase shift that can be obtained throughout the whole spectrum. The results calculated for a range of input powers are shown on the Fig. \ref{fig:ps2} a). 
As expected, the power dependence is linear due to the $|E|^2 \sim P$ factor in Eq. (\ref{deltafi}). However, the dependence on thickness, shown on the Fig. \ref{fig:ps2} b), is more complicated. Initially, as $L$ increases, the phase shift is also rapidly increasing, starting from $\Delta \Phi (L \rightarrow 0)=0$. However, at some point, the increase of the optical length is compensated by the decrease of $\chi^{(3)}$, which is enhanced in very thin wells. Thus, the phase shift reaches a local maximum and then starts decreasing with increasing $L$. Eventually, in the region of $L \sim 300$ nm, the value of $\chi^{(3)}$ stabilizes on the same level as in bulk medium and the phase shift again becomes linearly dependent on optical length. One can also notice slight oscillations on the Fig. \ref{fig:ps2} b) in the region $L \sim 100$ nm. In this regime, the confinement states are  mostly visible; the thickness-dependent overlapping of multiple states slightly affects the maximum value of the susceptibility and thus the phase shift. In conclusion, the choice of quantum well thickness, input power and specific excitonic state to realize a self-Kerr shift is highly nontrivial, with multiple tradeoffs influenced by amplification of nonlinear properties, overlap of confinement states and Rydberg blockade.

\section{Conclusions}
In summary, we have studied the nonlinear interaction between an electromagnetic wave and Rydberg excitons in Cu$_2$O quantum well using the Real Density Matrix Approach, incorporating the control  of the Rydberg blockade. Our theoretical, analytical results for linear and nonlinear absorption are illustrated by numerical calculations  and indicate the potential experimental conditions for the best observation of confinement states in linear and nonlinear optical spectra.  We show that a clear separation of confinement states and an amplification of nonlinear properties of the system are possible in sufficiently thin ($L < 100$ nm) quantum wells. The interplay between nonlinearity enhancement and optical length of the system is discussed. 
We theoretically demonstrate that the Kerr nonlinearity and significant self-phase modulation are  accomplished in  a semiconductor quantum well with REs.
 \\\noindent In short, our work provides insights into the nonlinear interactions of RE with photons  in quantum-confined systems, opening  interesting opportunities to explore Rydberg excitons for future opto-electronic nanoscale applications. We hope that our results might be useful  for future direct integration of Rydberg confined states with nanophotonic devices.

\appendix
\begin{small}
\section{Coefficients A,B}\label{App1}
\noindent Denoting by $\tilde{f}_{0e}$ and $\tilde{f}_{0h}$ the modified Boltzmann distributions $f_{0e}$ and $f_{)h}$ for electron and holes respectively, projections of $f_{0e,h}$ on the eigenfunctions $\Psi_{\ell
1N_eN_h}$ are given by the following expressions
\begin{eqnarray*}
 &&\langle \Psi_{\ell
N_eN_h}\vert\tilde{f}_{0e}(\textbf{r})\rangle=A_{\ell N_eN_h},\\
&&\langle \Psi_{\ell
N_eN_h}\vert\tilde{f}_{0h}(\textbf{r})\rangle=B_{\ell N_eN_h},\\
\end{eqnarray*}
which can be used to calculate the constants
\begin{eqnarray*}
&&A_{\ell N_eN_h}=\langle
\psi_{j}(r,\phi)\vert\tilde{f}^\parl_{0e}(r,\phi)\rangle\langle
\Psi_{N_eN_h}\vert\tilde{f}_{0e}^\perp(z_e,z_h)\rangle,\\
&&=\sqrt{\frac{\pi}{2}}\frac{1}{\kappa_{\ell}^{1/2}}{\mathcal
A}\,(2\kappa_{\ell})^\beta\,M\left(-\ell,3,4\kappa_{\ell
1}\rho_0\right)\sqrt{(\ell+1)(\ell+2)}\, I_{N_eN_h}^{(e)},\\
 &&B_{\ell N_eN_h}=\langle
\psi_{j}(r,\phi)\vert\tilde{f}^\parl_{0h}(r,\phi)\rangle\langle
\Psi_{N_eN_h}\vert\tilde{f}_{0h}^\perp(z_e,z_h)\rangle\\
&&=\sqrt{\frac{\pi}{2}}\frac{1}{\kappa_{\ell}^{1/2}}{\mathcal
B}\,(2\kappa_{\ell})^\gamma\,M\left(-\ell,3,4\kappa_{\ell
}\rho_0\right)\sqrt{(\ell+1)(\ell+2)}\, I_{N_eN_h}^{(h)},
\end{eqnarray*}
where the following approximation has been used
\begin{eqnarray*}
&&\langle
\psi_{j}(r,\phi)\vert\tilde{f}^\parl_{0e}(r,\phi)\rangle\\
&&=\int\limits_0^\infty \rho\,d\rho\, R_{\ell
1}(\rho)\sqrt{\frac{\pi}{2}}\frac{\rho}{\tilde{\lambda}_{\hbox{\tiny
th e}}}\exp\left[-\frac{\rho^2}{2\tilde{\lambda}_{\hbox{\tiny th
e}}^2}\right]\\
&&=\int\limits_0^\infty
\rho\,d\rho\,\sqrt{\frac{\pi}{2}}\frac{\rho}{\tilde{\lambda}_{\hbox{\tiny
th e}}}\exp\left[-\frac{\rho^2}{2\tilde{\lambda}_{\hbox{\tiny th
e}}^2}\right]\\
&&\times C_{\ell}(4\kappa_{\ell}\rho)e^{-2\kappa_{\ell}\rho}
M\left(-\ell,3,4\kappa_{\ell}\rho\right)\\
&&\approx \frac{4\kappa_{\ell}}{\tilde{\lambda}_{\hbox{\tiny th
e}}}\sqrt{\frac{\pi}{2}}C_{\ell}M\left(-\ell,3,4\kappa_{\ell}\rho_0\right)\int\limits_0^\infty
\rho^3\,d\rho\,e^{-2\kappa_{\ell}\rho-\frac{\rho^2}{2\tilde{\lambda}_{\hbox{\tiny th e}}^2}}\\
&&=\frac{1}{4\kappa_{\ell}^3{\tilde{\lambda}_{\hbox{\tiny the}}}}\sqrt{\frac{\pi}{2}}C_{\ell}M\left(-\ell,3,4\kappa_{\ell}\rho_0\right)\exp[f(x,z)]\\
&&=\frac{1}{2\kappa_{\ell}^{3/2}{\tilde{\lambda}_{\hbox{\tiny th e}}}}\sqrt{\frac{\pi}{2}}\sqrt{(\ell+1)(\ell+2)}M\left(-\ell,3,4\kappa_{\ell}\rho_0\right)\exp[f(x,z)]\\
 &&f(x,z)=4\ln z_\ell^{(e)}+3\ln x_{\ell}^{(e)}+\ln
[\sigma_{\ell}^{(e)}\sqrt{2\pi}]-\frac{1}{2}x_{\ell}^{(e)2}-z_\ell^{(e)}x_\ell^{(e)},
\end{eqnarray*}
and the integral is evaluated as follows
\begin{eqnarray*}
&&\int\limits_0^\infty \rho^3\,d\rho\,e^{-2\kappa_{\ell}\rho-\frac{\rho^2}{2\tilde{\lambda}_{\hbox{\tiny th e}}^2}}\\
&&=\frac{1}{(2\kappa_{\ell})^4}(z^{(e)}_{\ell})^4\Gamma(4)e^{(z^{(e)}_{\ell})^2/4}D_{-4}(z^{(e)}_{\ell}),\\
&&z^{(e)}_{\ell}=2\kappa_{\ell}\tilde{\lambda}_{\hbox{\tiny th
e}}=\frac{2\tilde{\lambda}_{\hbox{\tiny th e}}}{2\ell+3}.
\end{eqnarray*}
The $D_{-4}\left[z^{(e)}_{\ell}\right]$ is the parabolic
cylinder function \cite{Abramovitz}, and
$$z^{(e)}_{\ell}=2\kappa_{\ell}\tilde{\lambda}_{\hbox{\tiny th
e}}=\frac{2\tilde{\lambda}_{\hbox{\tiny th e}}}{2\ell+3},\quad
\tilde{\lambda}_{\hbox{\tiny th e}}=\frac{{\lambda}_{\hbox{\tiny
th e}}}{a^*}.$$ The term containing function $D_\nu$ can be
approximated as follows
\begin{eqnarray*}
&&z_\ell^{(e)4}\Gamma(4)e^{z_\ell^{(e)2}/4}D_{-4}\left[z_\ell^{(e)}\right]\approx z_\ell^{(e)4}\sigma^{(e)}_{\ell}\sqrt{2\pi}\exp\left\{f\left[x_\ell^{(e)}\right]\right\},\\
&&x_\ell^{(e)}=\frac{\sqrt{12+z_\ell^{(e)2}}-z_\ell^{(e)}}{2},\\
&&\sigma^{(e)}_{\ell}=\left(1+\frac{3}{x_\ell^{(e)2}}\right)^{-1/2},\\
&&f\left[x_\ell^{(e)}\right]=3\ln x_\ell^{(e)}-\frac{1}{2}x_\ell^{(e)2}-z_\ell^{(e)}x_\ell^{(e)}.\\
\end{eqnarray*}

\section{calculation of $\chi^{(3)}$}\label{App2}
The equation (\ref{chi3}) can be written in the form
\begin{eqnarray*}\label{chi32}
&&\chi^{(3)}(\omega)=-\epsilon_0(\epsilon_b\Delta_{LT})^2a^{*3}e^{-4\rho_0}\left(\frac{1}{{\mit\Gamma}_{01}}\right)\nonumber\\
&&\times\sum\limits_{jn_en_h}{\mit\Gamma_j}\langle\Psi_{N_eN_h}\rangle_L\frac{\Psi_{jn_en_h}(r_0)\sqrt{f_{j}}}{(E_{Tjn_e n_h}-\hbar\omega)^2+{\mit\Gamma_j}^2}\\
&&\times\sum\limits_{\ell N_eN_h}\frac{\sqrt{f_{\ell}}E_{T1\ell
N_eN_h}(A_{\ell N_eN_h}+B_{\ell N_eN_h})}{E_{T\ell N_e
N_h}^2-(\hbar\omega+i{\mit\Gamma}_{\ell
N_eN_h})^2},\end{eqnarray*} where
\begin{eqnarray*}
&&\Psi_{\ell N_eN_h}=\psi_{\ell
}(r,\phi)\psi_{L,N_e}^{(1D)}(z_e)\psi_{L,N_h}^{(1D)}(z_h),\\
&&\psi_{\ell m}(r,\phi)=R_{\ell m}\frac{e^{im\phi}}{\sqrt{2\pi}}=\frac{1}{a^*}\frac{e^{im\phi}}{\sqrt{2\pi}}\nonumber\\
&&\times
e^{-2\kappa_{\ell m}r/a^*}\left(4\kappa_{\ell m}\frac{r}{a^*}\right)^m\,4\kappa_{\ell m}^{3/2}\frac{1}{(2m)!}\frac{[(\ell+2m)!]^{1/2}}{[\ell!]^{1/2}}\\
&&\times M\left(-\ell,2\vert m\vert+1,4\kappa_{\ell m}\frac{r}{a^*}\right),\nonumber\\
&&=\Phi_m(\phi) C_{\ell m}\left(4\kappa_{\ell
m}\frac{r}{a^*}\right)^m e^{-2\kappa_{\ell
m}r/a^*}\,M\left(-\ell,2\vert
m\vert+1,4\kappa_{\ell m}\frac{r}{a^*}\right),\nonumber\\
&&\kappa_{\ell m}=\frac{1}{1+2(\ell+\vert m\vert)},\nonumber\\
&&C_{\ell m}=\frac{1}{a^*}4\kappa_{\ell m}^{3/2}\frac{1}{(2m)!}\frac{[(\ell+2m)!]^{1/2}}{[\ell!]^{1/2}},\\
&&\psi_{L,N_e}^{(1D)}(z_e)=\sqrt{\frac{2}{L}}\cos\left[(2N_e-1)\pi\frac{z_e}{L}\right],\\
&&\psi_{L,N_h}^{(1D)}(z_h)=\sqrt{\frac{2}{L}}\cos\left[(2N_h-1)\pi\frac{z_h}{L}\right],\\
 &&A_{\ell N_eN_h}=\langle \Psi_{\ell
N_eN_h}\vert\tilde{f}_{0e}(\textbf{r})\rangle,\\ &&A_{\ell
N_eN_h}=\langle \psi_{\ell}(r,\phi)\vert\tilde{f}^\parl_{0e}(r,\phi)\rangle\langle
\Psi_{N_eN_h}\vert\tilde{f}_{0e}^\perp(z_e,z_h)\rangle,\\
&&B_{\ell N_eN_h}=\langle \psi_{\ell}(r,\phi)\vert\tilde{f}^\parl_{0h}(r,\phi)\rangle\langle
\Psi_{N_eN_h}\vert\tilde{f}_{0h}^\perp(z_e,z_h)\rangle,\\
&&\Psi_{N_eN_h}=\psi_{L,N_e}^{(1D)}(z_e)\psi_{L,N_h}^{(1D)}(z_h),
\end{eqnarray*}
\begin{eqnarray}\label{edistribution_2}
&&\tilde{f}_{0e}(\textbf{r})=\tilde{f_{0e}}(\rho,z_e,z_h,\phi)=\sqrt{\frac{\pi}{2}}\frac{r}{\lambda_{\hbox{\tiny
th e}}}\nonumber\\
&&\times
[\Phi_1(\phi)+\Phi_{-1}(\phi)]\exp\left(-\frac{r^2+(z_e-z_h)^2}{2}\frac{m_e
k_B{\mathcal T}}{\hbar^2}\right)\nonumber\\
&&=\tilde{f}_{0e}^\perp(z_e,z_h)\tilde{f}^\parl_{0e}(r,\phi),\nonumber\\
&&\tilde{f}_{0e}^\perp(z_e,z_h)=\exp\left(-\frac{(z_e-z_h)^2}{2}\frac{m_e
k_B{\mathcal T}}{\hbar^2}\right),\\
&&\tilde{f}^\parl_{0e}(r,\phi)=\sqrt{\frac{\pi}{2}}\frac{r}{\lambda_{\hbox{\tiny
th e}}}\exp\left(-\frac{r^2}{2}\frac{m_e k_B{\mathcal
T}}{\hbar^2}\right)[\Phi_1(\phi)+\Phi_{-1}(\phi)],\nonumber\\
 &&r=\sqrt{x^2+y^2},\nonumber
\end{eqnarray}
and
\begin{eqnarray*}
&&\Phi_m(\phi)=\frac{e^{im\phi}}{\sqrt{2\pi}},
\end{eqnarray*}
 \begin{eqnarray*}&&\lambda_{\hbox{\tiny th
e}}=\left(\frac{\hbar^2}{m_e k_B{\mathcal
T}}\right)^{1/2}=\sqrt{\frac{2\mu}{m_e}}\sqrt{\frac{R^*}{k_B{\mathcal
T}}}a^*,\end{eqnarray*} is the so-called thermal length (here for
electrons).

Similarly, for the hole equilibrium distribution, we have
\begin{eqnarray}\label{edistribution3}
&&\tilde{f}_{0h}(\textbf{r})=\tilde{f_{0h}}(\rho,z_e,z_h,\phi)=\sqrt{\frac{\pi}{2}}\frac{r}{\lambda_{\hbox{\tiny
th h}}}\nonumber\\
&&\times
[\Phi_1(\phi)+\Phi_{-1}(\phi)]\exp\left(-\frac{r^2+(z_e-z_h)^2}{2}\frac{m_h
k_B{\mathcal T}}{\hbar^2}\right)=\nonumber\\
&&=\tilde{f}_{0h}^\perp(z_e,z_h)\tilde{f}^\parl_{0h}(r,\phi),\\
&&\tilde{f}_{0h}^\perp(z_e,z_h)=\exp\left(-\frac{(z_e-z_h)^2}{2}\frac{m_h
k_B{\mathcal T}}{\hbar^2}\right),\nonumber\\
&&\tilde{f}^\parl_{0h}(r,\phi)=\sqrt{\frac{\pi}{2}}\frac{r}{\lambda_{\hbox{\tiny
th h}}}\exp\left(-\frac{r^2}{2}\frac{m_h k_B{\mathcal
T}}{\hbar^2}\right)[\Phi_1(\phi)+\Phi_{-1}(\phi)],\nonumber
\end{eqnarray}
\noindent with the hole thermal length
\begin{eqnarray*}&&\lambda_{\hbox{\tiny th
h}}=\left(\frac{\hbar^2}{m_h k_B{\mathcal
T}}\right)^{1/2}=\sqrt{\frac{2\mu}{m_h}}\sqrt{\frac{R^*}{k_B{\mathcal
T}}}a^*,\end{eqnarray*} 
with the radial part $\psi_{j}(r,\phi)\vert\tilde{f}^\parl_{0e}(r,\phi)\rangle$ defined in Appendix \ref{App1}.

\section{table of parameters}\label{App3}
\begin{table}[ht!]
\caption{\small Band parameter values for Cu$_2$O, masses in free
electron mass $m_0$.}
\begin{center}
\begin{tabular}{p{.2\linewidth} p{.2\linewidth} p{.2\linewidth} p{.2\linewidth} p{.2\linewidth}}
\hline
Parameter & Value &Unit&Reference\\
\hline $E_g$ & 2172.08& meV& \cite{Kazimierczuk2014}\\
$R^*$&87.78& meV &\cite{FK}\\
$\Delta_{LT}$&$1.25\times 10^{-3}$&{meV}& \cite{Stolz}\\
$m_e$ & 0.99& $m_0$&\cite{Naka}\\
$m_h$ &0.58&  $m_0$&\cite{Naka}\\
$\mu$ & 0.363 &$m_0$&\\
$\mu'$&-2.33&$m_0$&\\
$M_{tot}$&1.56& $m_0$&\\
$a^*$&1.1& nm&\cite{FK}\\
$r_0$&0.22& nm&\cite{Zielinska.PRB}\\
$\epsilon_b$&7.5 &&\cite{Kazimierczuk2014}\\
$T_1$&500&ns&\\ \hline
\end{tabular} \label{parametervalues}\end{center}
\end{table}
\end{small}


\begin{thebibliography}{99}
\bibitem {Kazimierczuk2014}
T. Kazimierczuk, D. Fr\"{o}hlich, S. Scheel, H. Stolz, and M. Bayer, Nature \textbf{514}, 344 (2014).

\bibitem{Heckotter2017}
J. Heck\"{o}tter, M. Freitag, D. Fr\"{o}hlich, M. A{\ss}mann, M. Bayer, M. A. Semina, and M. M. Glazov, Phys. Rev.B \textbf{95}, 035210 (2017).

\bibitem{Assman2020}
M. Assmann, and M. Bayer, Adv. Quantum Technol. \textbf{3}, 1900134 (2020). 

\bibitem{Lynch2021}
S. A. Lynch, C. Hodges, S. Mandal, W. Langbein, R. P. Singh, L. Gallagher, J. D. Pritchett, D. Pizzey, J. P. Rogers, C. Adams, and M. P. Jones, Phys. Rev. Materials \textbf{5}, 084602 (2021).

\bibitem{Heckotter2021}
J. Heckötter, V. Walther, S. Scheel, M. Bayer, T. Pohl, and M. Assmann, Nature Communications \textbf{12}, 3556 (2021).

\bibitem{Liam2022}
L.A.P. Gallagher, J.P. Rogers, J.D. Pritchett, R.A. Mistry, D. Pizzey, Ch.S. Adams, M.P.A. Jones, P. Grünwald, V. Walther, Ch. Hodges, and W. Langbein, and S. A. Lynch, Phys. Rev. Research \textbf{4}, 013031 (2022).

\bibitem{Orfanakis2022}
K. Orfanakis, S. Rajendran, V. Walther, T. Volz, T. Pohl, and H. Ohadi, Nature Materials (2022).

\bibitem{Raczynska2019}
S. Zieli\'{n}ska-Raczy\'{n}ska, G. Czajkowski, K. Karpi\'{n}ski, and D. Ziemkiewicz, Phys. Rev. B \textbf{99}, 245206 (2019).

\bibitem{Walther2020}
V. Walther, P. Gr\"unwald, and T. Pohl, Phys. Rev. Lett. \textbf{125}, 173601 (2020).

\bibitem{ThomasArxiv2022}
C. Morin, J. Tignon, J. Mangeney, S. Dhillon, G. Czajkowski, K. Karpiński, S. Zielińska-Raczyńska, and D. Ziemkiewicz, T. Boulier, arXiv:2202.09239v1 [quant-ph].

\bibitem{Walther2018}
V. Walther, R. Johne, and T. Pohl, Nature Communications \textbf{9}, 1309 (2018).

\bibitem{Ziemkiewicz2018}
D. Ziemkiewicz, and S. Zielińska-Raczyńska, Optics Letters \textbf{43}, 3742-3745 (2018).

\bibitem{Ziemkiewicz2019}
D. Ziemkiewicz, and S. Zielińska-Raczyńska, Optics Express \textbf{27(12)}, 16983 (2019).

\bibitem{Konzelmann2019}
A. Konzelmann, B. Frank, and H. Giessen, J. Phys. B \textbf{53}, 024001 (2020). 

\bibitem{Czajkowski2020}
D. Ziemkiewicz, K. Karpiński, G. Czajkowski, and S. Zielińska-Raczyńska, Phys. Rev. B \textbf{101}, 205202 (2020).

\bibitem{Czajkowski2021}
D. Ziemkiewicz, G. Czajkowski, K. Karpiński, and S. Zielińska – Raczyńska, Phys. Rev. B \textbf{103}, 035305 (2021).

\bibitem{Orfanakis2021}
K. Orfanakis, S. Rajendran, H. Ohadi, S. Zielińska-Raczyńska, G. Czajkowski, K. Karpiński, and D. Ziemkiewicz, Phys. Rev. B \textbf{103}, 245426 (2021).

\bibitem{Takahata}
M. Takahata, K. Tanaka, and N. Naka, Phys. Rev. B \textbf{97}, 205305, (2018)

\bibitem{Steinhauer}
S. Steinhauer, M.A.M. Versteegh, S. Gyger, A. Elshaari, B. Kunert, A. Mysyrowicz, and V. Zwiller,
 Commun Mater 1, 11 (2020). https://doi.org/10.1038/s43246-020-0013-6

\bibitem{Hamedi}
H. R. Hamedi, and M. R. Mehmannavaz, Physica E: Low-dimensional Systems and Nanostructures \textbf{66}, 309-316, (2015).

\bibitem{Kosionis}
S. G. Kosionis, A. F. Terzis, and E. Paspalakis, Journal of Applied Physics \textbf{109(8)}, 084312 (2011).

\bibitem{Zhu}
C. Zhu, and G. Huang, Opt Express \textbf{19(23)}, 23364-23376 (2011).

\bibitem{Qian}
H. Qian, Y. Xiao, and Z. Liu, Nat Commun \textbf{7}, 13153 (2016).

\bibitem{StB87}
A. Stahl and I. Balslev, {\sl Electrodynamics of the Semiconductor
Band Edge} (Springer-Verlag, Berlin-Heidelberg-New York, 1987).

\bibitem{CBass}
{G. Czajkowski, F. Bassani, and A. Tredicucci}, Polaritonic
effects in superlattices, Phys. Rev. B \textbf{54}, 2035 (1996).

\bibitem{PRB93}
S. Zielińska – Raczyńska, G. Czajkowski, and D. Ziemkiewicz, Phys. Rev. B \textbf{93}, 075206 (2016).

\bibitem{Abramovitz}
M. Abramowitz and I. Stegun, \emph{Handbook of Mathematical Functions} (Dover Publications, New York, 1965).

\bibitem{Kang}
D. Kang, A. Gross, H. Yang, Y. Morita, K. Choi, K. Yoshioka, and N. Y. Kim, Phys Rev. B \textbf{103}, 205203 (2021).

\bibitem{FrankStahl}
D.~Frank and A.~Stahl, Solid State Commun. {\bf 52}, 861 (1984).

\bibitem{Heckotter2018}
J. Heck\"{o}tter, M. Freitag, D. Fr\"{o}hlich, M. Assmann, M. Bayer, P. Gr\"{u}nwald, F. Sch\"{o}ne, D. Semkat, H. Stolz, and S. Scheel, Phys. Rev. Lett. \textbf{121}, 097401 (2018).

\bibitem{FK}
Zieli\'{n}ska-Raczy\'{n}ska, D. Ziemkiewicz, and G. Czajkowski, Phys. Rev. B \textbf{97}, 165205 (2018).

\bibitem{Stolz}
H. Stolz,  F. Sch\"{o}ne, and D. Semkat, New. J. Phys. \textbf{20}, 023019, (2018).

\bibitem{Naka}
{N. Naka, I. Akimoto, M. Shirai, and  Ken-ichi Kan'no}, Phys. Rev. B \textbf{85}, 035209 (2012).

\bibitem{Zielinska.PRB}
S. Zieli\'{n}ska-Raczy\'{n}ska, G. Czajkowski, and D. Ziemkiewicz, Phys. Rev. B \textbf{93}, 075206 (2016).
\end{thebibliography}
\end{document}